\documentclass[useAMS,usenatbib,usegraphicx]{mn2e}
 
\title[The GOODS NICMOS Survey]{The Hubble Space Telescope GOODS NICMOS Survey: Overview and the Evolution of Massive Galaxies at $1.5 < z < 3$}
\author[Christopher J. Conselice et al. ]{C.J. Conselice$^{1}$\thanks{E-mail:
conselice@nottingham.ac.uk},  A.F.L. Bluck$^{1}$, F. Buitrago$^{1}$, A.E. Bauer$^{1}$, R. Gr\"utzbauch$^{1}$, \newauthor R.J. Bouwens$^{2,3}$, S. Bevan$^{1}$, A. Mortlock$^{1}$, M. Dickinson$^{4}$, E. Daddi$^{5}$, H. Yan$^{6}$,  \newauthor Douglas Scott$^{7}$, S.C. Chapman$^{8}$, R.-R. Chary$^{9}$, H.C. Ferguson$^{10}$, M. Giavalisco$^{11}$, \newauthor N. Grogin$^{10}$, G. Illingworth$^{2}$, S. Jogee$^{12}$, A.M. Koekemoer$^{10}$, Ray A. Lucas$^{10}$, \newauthor B. Mobasher$^{13}$, L. Moustakas$^{14}$, C. Papovich$^{15}$, S. Ravindranath$^{16}$,  B. Siana$^{17}$,  \newauthor H. Teplitz$^{18}$, I. Trujillo$^{19}$, M. Urry$^{20}$, T. Weinzirl$^{12}$ \\
\\
\\
$^{1}$University of Nottingham, School of Physics \& Astronomy, Nottingham, NG7 2RD UK \\
$^{2}$Lick Observatory, University of California, Santa Cruz \\
$^{3}$Leiden Observatory, Leiden University, NL-2300 RA Leiden, Netherlands \\
$^{4}$National Optical Astronomical Observatories, Tucson, AZ \\
$^{5}$CEA/DSM-CNRS, CEA Saclay \\
$^{6}$Ohio State University, Columbus, Ohio \\
$^{7}$Department of Physics, University of British Columbia \\
$^{8}$Institute of Astronomy, Cambridge University, UK \\
$^{9}$Spitzer Science Center, California Institute of Technology \\
$^{10}$Space Telescope Science Institute, Baltimore, MD \\
$^{11}$Department of Astronomy, University of Massachuetts, Amherst \\
$^{12}$Department of Astronomy, University of Texas, Austin \\
$^{13}$Department of Physics, University of California, Riverside \\
$^{14}$Jet Propulsion Laboratory, California Institute of Technology \\
$^{15}$Department of Physics \& Astronomy, Texas A\&M University \\
$^{16}$IUCAA, Pune University Campus, Pune 411007, Maharashtra, India \\
$^{17}$Astronomy, Caltech, MC 105-24, Pasadena 91125 \\
$^{18}$Infrared Processing and Analysis Center, Caltech \\
$^{19}$IAC, Spain \\
$^{20}$Department of Physics, Yale University }

\def\solm{M$_{\odot}\,$}

\def\solm{M$_{\odot}\,$}

\def\mass{$10^{11}$$M_{\odot}\,$}
\def\hmass{$10^{11.5}$$M_{\odot}\,$}
\def\lmass{$10^{10}$$M_{\odot}\,$}

\def\casgm20{CAS-G-M$_{20}\,$}
\def\m20{M$_{20}\,$}
\begin{document}

\date{Accepted ; Received ; in original form}
\pagerange{\pageref{firstpage}--\pageref{lastpage}} \pubyear{2002}

\maketitle

\label{firstpage}
\clearpage
\begin{abstract}

We present the details and  early results from a deep near-Infrared survey 
utilising the NICMOS instrument on the {\it Hubble Space Telescope} 
centred around
massive $M_{*} > 10^{11}$ \solm galaxies at $1.7 < z < 2.9 $ found within the 
Great Observatories
Origins Deep Survey (GOODS) fields North and South. The GOODS NICMOS Survey 
(GNS) was designed to obtain deep F160W
(H-band) imaging of 80 of these massive galaxies, as well as other colour
selected objects
such as Lyman-break drop-outs, BzK objects, Distant Red Galaxies (DRGs), 
Extremely Red Objects (EROs), {\it Spitzer} Selected
EROs, BX/BM galaxies, as well as flux selected sub-mm galaxies.  
We present in this paper details of the observations, our sample
selection, as well as a description of features of the massive
galaxies found within our survey fields. This includes: photometric redshifts, 
rest-frame colours, and  stellar masses.  We furthermore provide an 
analysis of the
selection methods for finding massive galaxies at high redshifts,
including colour selection methods and how galaxy populations selected
through these colour methods overlap.
We find that a single colour selection method cannot locate all of the massive
galaxies, with no one method finding more than 70 percent.  
We however find that the combination of these colour methods finds
nearly all the massive galaxies, as selected by photometric redshifts with
the exception of apparently rare blue massive galaxies.  By investigating the 
rest-frame $(U-B)$ vs. M$_{\rm B}$ diagram for these galaxies we furthermore show
that there exists a bimodality in colour-magnitude space at $z < 2$, driven by
stellar mass, such that the most massive galaxies are systematically red up to 
$z \sim 2.5$, while lower mass galaxies tend to be blue.  We also
discuss the number densities for galaxies with stellar masses 
$M_{*} > 10^{11}$ \solm, whereby we find an increase of a factor of eight
between $z = 3$ and $z = 1.5$, demonstrating that this is an epoch when massive
galaxies establish most of their mass.   We also provide an overview of the
evolutionary properties of these galaxies, such as their merger histories, 
and size evolution.  

\end{abstract}

\section{Introduction}

Our understanding of distant galaxies and the history of galaxy formation has 
undergone a revolution in the past decade. Galaxies are now routinely
discovered and studied out to redshifts $z \sim 4-6$ (e.g., Dickinson
et al. 2004;  Yan
et al. 2005; Bouwens et al. 2007; Bouwens et al. 2010).    
Samples of a few dozen objects have been 
found at even higher redshift, back to the era  of reionization 
($z \sim 6-7$),  and perhaps some galaxies have been discovered 
at even higher redshifts, $z \sim 8-10$ (e.g., Bouwens et al. 2010;  
Finkelstein et al.  2010).   This relatively rapid advance in our 
discovery of the earliest galaxies is the direct result of technical
advances in spectroscopy and imaging over the past decade, in which the
{\it Hubble Space Telescope} ({\it HST}) has played a leading role. 

Historically, distant galaxies are found within deep optical imaging surveys, 
and are confirmed as  high redshift galaxies with large multi-object 
spectrographs on 8-10 meter telescopes, which came online in the mid-1990s.  
It can be argued however that some of the most important advances in our
understanding of galaxies has come about from very deep imaging, especially
from {\it HST}. The {\it Hubble} 
Space Telescope has played a  key role in high-redshift
discoveries and our understanding of galaxy evolution through large blank 
field and 
targeted programs such as the {\it Hubble} Deep and 
Ultra Deep Fields, GOODS, EGS, and COSMOS, among others 
(e.g., Williams et al. 1996; Giavalisco et al. 2004; Beckwith et al. 2006; 
Davis et al. 2007; Scoville et al. 2007).

This {\it Hubble} imaging has proven invaluable for two primary reasons.  One
is simply due to the depth that can be achieved with a high photometric
fidelity, ensuring that exquisite photometry of distant galaxies can be
obtained.  Whilst ground based telescopes can reach the depths of {\it HST} 
at optical wavelengths, in principle the accuracy 
and precision of this photometry is
not nearly as good, due to a higher background, and importantly, the
large and variable PSF. This makes accurate measurements of light difficult, 
particularly for colours
which require exact apertures for accurate measures.  Furthermore, 
{\it HST} data have proven important for the discovery of the most distant galaxies
in the Universe through the use of the Lyman-Break method of looking for
drop-out galaxies in bluer bands.  Many filter choices within multi-colour 
deep imaging programmes were in fact selected to facilitate optimal 
drop-out searches (e.g., Giavalisco et al. 2004). 
 
{\it Hubble} imaging furthermore has facilitated a renaissance in the study of
galaxy structure in the distant Universe, which provides a key observable
for understanding how distant galaxies form and evolve (e.g., Conselice
et al. 2003; Conselice et al. 2008, 2009; Lotz et al. 2008; Buitrago
et al. 2008; Jogee et al. 2009; Bluck et al. 2009; 
Casatta et al. 2010).   These
structural measurements have proven critical for determining how galaxy
morphologies, sizes, and merger/kinetic states have evolved through 
time (e.g., Trujillo et al. 2007; Ravindranath et al. 2004; Conselice
et al. 2003).  This allows us to examine how the merger history
of galaxies has changed (e.g., Conselice et al. 2003, 2008; Lotz
et al. 2008; Jogee et al. 2009), and thus we can  begin to derive 
how galaxies form, as opposed
to simply when.  It is not currently straightforward to measure the structures
of distant galaxies with ground based imaging even with adaptive optics, 
and thus {\it Hubble} has and continues to provide a key aspect for tracing 
evolution using these methods.

However, one key aspect of parameter
space that has not yet been explored with {\it HST}, or other space-based
telescopes in any depth over large areas, is deep infrared imaging 
over a relatively large area.  Previously there exists deep NIC3 imaging
over the {\it Hubble} Deep Field (Dickinson et al. 2000) and Hubble
Ultra-Deep Field (Thompson et al. 2004), as well
as very deep NICMOS imaging over a small area of the HDF-N (Thompson 
et al. 1999).   These areas are however very small, and while NIC3 parallel
data exists over the COSMOS and EGS fields, it is quite shallow at $\sim
1$ orbit depth.    {\it HST} imaging data however has a distinct 
advantage over 
ground-based imaging not only in terms of the higher quality photometric
fidelity and higher resolution, but also the depth which can be
achieved in the near infrared (NIR) with {\it HST} -- as opposed to the 
ground-based optical where comparable depths to {\it HST} can be reached. 
Within one or two orbits, the {\it HST} can reach a depth in the NIR 
which is difficult to obtain from the ground even with an 8-10-m class
telescope, and which will
not have the same photometric quality, nor resolution as the {\it Hubble}
data.

We thus designed and carried out the GOODS NICMOS Survey (GNS), which is a
large {\it HST} programme intended to remedy this situation by 
providing through an initial 180-orbit program of NIC3 imaging in
the GOODS fields, a data set designed to examine  a host of problems
requiring very deep NIR data.   The GNS data consist of 60 NICMOS
NIC3 pointings, centred on the most massive 
($M_\ast > 10^{11}$$M_\odot$) galaxies at $1.7 < z < 2.9$.  The depth of
each image is 3 orbits/pointing within the  $H_{160}$-bandpass 
over a total area of $\sim 43$ arcmin$^{2}$ (Buitrago et al. 2008;
Bluck et al. 2009, Casey et al. 2009, Bouwens et al. 2009, 2010 present 
results using this data).  

With these NICMOS data we are able to explore the rest-frame optical
features of galaxies at $z > 1$ in detail.  This allows a few
measurements to be made that cannot be easily reproduced 
with optical imagining 
and/or deep NIR imaging from the ground. This includes: filling in the 
important 
near-infrared gap in galaxy spectral energy distributions (SEDs);
sampling the rest-frame optical structures and sizes 
of galaxies out to $z \sim 3$ (Buitrago et al. 2008); and the
detection and characterization of the population of massive $z \sim 7-10$ 
galaxies and AGN, and determining the relation of AGN evolution to 
that of
massive galaxies (e.g., Bluck et al. 2010).

In this paper we present the basic outline, background, and results from this
survey. We discuss the design of the observations, our field selection,
as well as the selection for our initial massive galaxy sample which
has guided the centres for our NICMOS NIC3 pointings.  We also discuss
the various methods for locating the massive galaxy population at
higher redshifts, and the connection of these massive galaxies to those
at $z < 2$.  We show that no one colour method is able to identify the
massive galaxy population at high redshifts, and that a combination of
methods and photometric redshifts are needed to construct a semi-complete
massive galaxy sample at higher redshifts. In this paper we construct
as complete as possible sample of massive galaxies within our fields, 
and discuss
the properties of these galaxies, as well as some features of lower
mass galaxies.

This paper is organised as follows: \S 2 gives a summary of our observations
and the design of the GNS, including how the initial sample of galaxies
was selected.  \S 3 gives a description of the derived parameters from
the $H_{160}-$band imaging, including photometric redshifts and catalogue matching.  
\S 4 describes our initial analysis of the survey data, including how the
various selections for massive galaxies at high redshifts compare,
while finally \S 5 is our summary.   
We use a standard cosmology of {\it H}$_{0} = 70$ km s$^{-1}$ Mpc$^{-1}$, and 
$\Omega_{\rm m} = 1 - \Omega_{\Lambda}$ = 0.3 throughout.

\section{Observations}

\subsection{Survey Design}

The GNS selection and field coverage is based on the previous optical 
ACS and ground-based imaging from
the original GOODS program (Giavalisco et al. 2004).
The GOODS programme
is a multi-wavelength campaign to obtain a coherent collection
of deep imaging and spectroscopy in two 150 arcmin$^{2}$ areas in the 
northern and southern hemispheres (GOODS-N and GOODS-S). 
These two fields are centred around the 
{\it Hubble} Deep Field-North (HDF-N) and {\it Chandra} Deep 
Field-South (CDF-S), 
which are areas of very low dust extinction, and minimal stellar and radio 
contamination. The existing GOODS/ACS fields match the coverage of 
the GOODS {\it Spitzer} program and cover the
2 Msec exposure of the {\it Chandra} Deep Field South and the 2 Msec exposure
of the {\it Chandra} Deep Field North (Luo et al. 2008). 
Large ongoing campaigns to obtain spectroscopy for the GOODS 
fields have also been carried out, including 3000 spectra as part of 
the  Keck Treasury Redshift Survey (Wirth et al. 2004).  Another $\sim 3000$
redshifts in GOODS-S have been measured from various ESO programs 
(e.g., Vanzella et al.\ 2008; Le Fevre et al.\ 2005; Popesso et al. 2009; Balestra et al. 2010).  

The co-moving volume probed by GOODS at high redshifts, $2 < z < 6$, is 
similar to the co-moving volume covered by the COSMOS field
 (e.g., Scoville et al. 2007) at $0.2 < z < 1$. Furthermore, due to its depth 
at all wavelengths the GOODS fields are thus an ideal
location for examining the formation and evolution of early galaxies.
Deep NIR imaging of these fields however is lacking, although
some deep NIR imaging has been obtained with ESO telescopes using
SOFI and ISAAC
for the GOODS-South, as well as deep CFHT WIRCAM imaging, Subaru imaging, 
and some Keck imaging
over the GOODS North (e.g., Kajisawa et al. 2009; Retzlaff et al. 2010; 
Wang et al. 2010). However, these data only reach modest depths of
$K_{\rm vega} \sim 22$ compared to our {\it HST} imaging. The depth 
of our suvey is only 
comparable to previous NICMOS deep programmes covering the HDF-North and 
HDF-South fields, as well as new near-IR data obtained with 
WFC3 (Cassata et al. 2010).

\begin{figure*}
 \vbox to 130mm{
\includegraphics[angle=0, width=180mm]{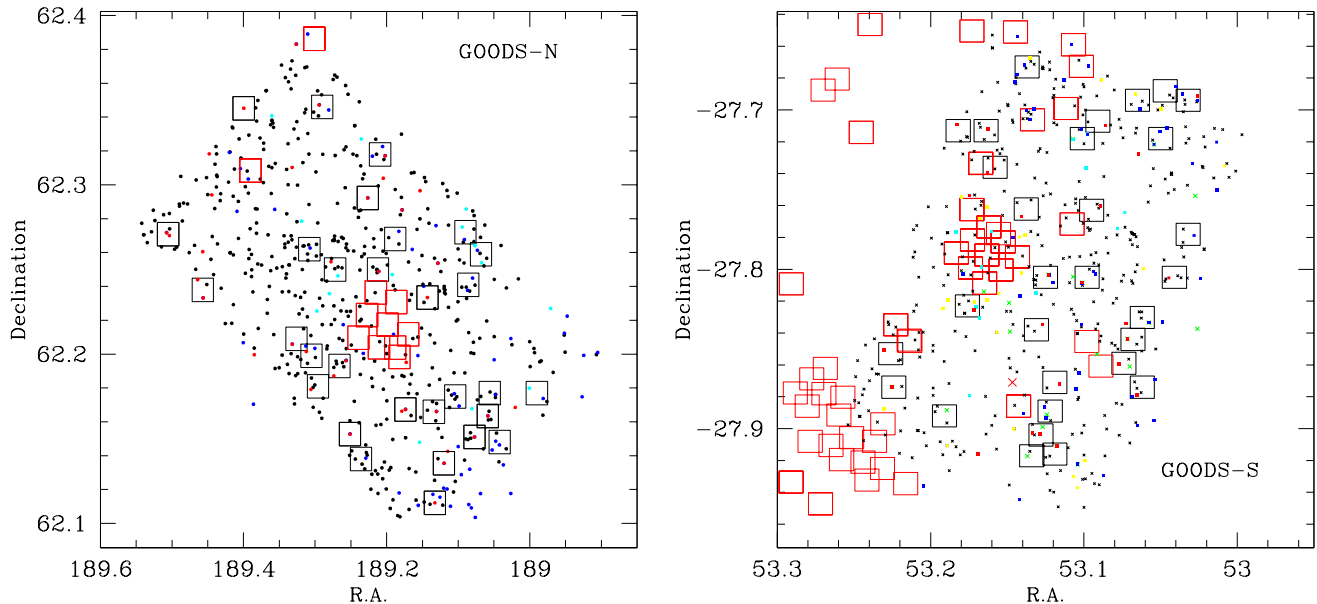}
 \caption{The distribution of the global targets used within the GOODS
North and South fields.  The points displayed are: DRGs (green), IEROs (blue),
BzKs (red), IRS {\it Spitzer} targets (cyan), V-drop outs (black), and high$-z$
galaxies with {\it Spitzer} IRS spectra (yellow).  The black boxes 
show the locations of the GNS
NICMOS pointings, while the red boxes are for previous deep NIC3 
fields in these areas.  The
{\it Hubble} Deep Field (within GOODS North) and the {\it Hubble} Ultra Deep
Field (within GOODS South) are shown in the centres of each field with
overlapping red tiles.  Note that these boxes have not been rotated to
match the orientation of these fields. }
} \label{sample-figure}
\end{figure*}

Ideally, one would want to cover both GOODS North fields completely, yet 
given the small NICMOS field of view it is not practical to cover
the entire GOODS fields any deeper than one orbit with NICMOS.  The WFC3
camera will, however, soon cover these fields to an even great depth with
the CANDELS programme.  Our 
strategy is not to map out a continuous area, but to collect 60 pointed 
observations directed towards the most massive galaxies at 
$z \sim 1.7 - 2.9$ found in the GOODS
fields (\S 2.2), maximised to obtain the largest number of galaxies 
based on our selection
methods.   To obtain the most
unique and useful science we therefore constructed a program which
covers a sixth of the area of a single GOODS field (in total
43.7 arcmin$^{2}$) in three orbits 
depth in the 
$H_{160}$ band, in areas of the deepest {\it Spitzer}, {\it Chandra}, and ACS
imaging, and where the greatest amount of spectroscopy already exists.  
Some of these fields were then observed in the $J$-band ($J_{110}$)
with NICMOS or WFC3 as
part of a follow up programme to obtain near infrared SEDs to look for 
high redshift drop-out galaxies (Bouwens et al. 2010).

\begin{figure*}
\hspace{4cm}
 \vbox to 130mm{
\includegraphics[angle=0, width=150mm]{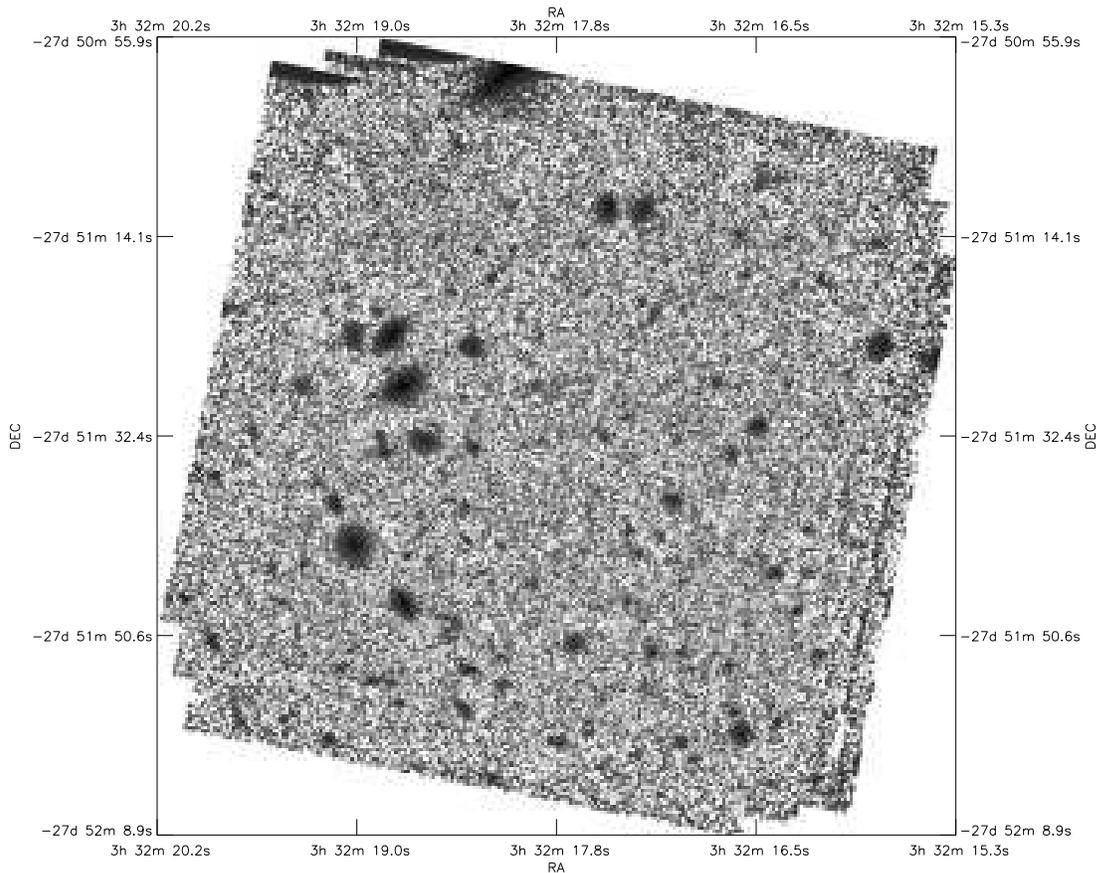}
 \caption{Example of one of our NICMOS images in the $H_{160}$ band.  This
example is for GOODS-South 16.  The field of view of this and all the NICMOS
pointings is approximately 51 arcsec on a side.}
} \label{sample-figure}
\end{figure*}

\subsection{Initial Galaxy Selection}

Our NICMOS pointings were chosen to target a set of objects selected to be 
known massive galaxies at high redshift, identified 
using a variety of color selection methods.  These include 
``Distant Red Galaxies'' (DRG: Franx et al. 2003;  Papovich et al. 2006), 
IRAC Extremely Red Objects (Yan et al. 2006), and
BzK color-selected galaxies (Daddi et al. 2004, 2007).  All of these
methods are designed to find red, dusty or passively evolving older
galaxies at $z > 1.5$.
In practice, we utilised all three of these colour-selections 
separately, in order to obtain as much as possible a complete sample 
of massive galaxies
at $z > 2$.  To optimise our field placement, we also used catalogues of
Lyman-break selected BM/BX objects (Reddy et al. 2008), as well
as high redshift drop-outs and sub-mm galaxies.  However, the primary
field selection was done in terms of the massive galaxy selection through
the three primary colour criteria as described further below.
 
Colour selection of distant galaxies has a long history dating back to the 
early work of finding Lyman-break galaxies through image drop-outs in blue
bands (e.g., Guhathakurta, Tyson, Majewski 1990; Steidel \& Hamilton 1992).  
It is generally accepted that no single method can find all galaxies at
a given redshift, and some of these methods are better at finding 
star-forming objects, as opposed to those which are more passive and 
evolved. In
fact, it is generally agreed that no method or combination of methods
can identify an obviously complete sample of high-z galaxies.

One of the methods we use for finding likely passively evolving and dusty
red galaxies 
is to find Distant Red Galaxies (DRGs) defined by a NIR colour cut (e.g.,
Franx et al. 2003; Papovich et al. 2006; Conselice et al. 2007b).  
The selection we use to 
find DRGs, and to be included within our sample, is galaxies at
$z \sim 1.7 - 2.9$ with ($J-K_{\rm s}$) $> 2.3$ mag in Vega magnitudes
 (or $> 1.37$ in AB mags).   The selection for
these galaxies is based on ground-based imaging from ISAAC on the 
Very Large Telescope (VLT).  This selection is only used 
for choosing systems in GOODS-S, as deep NIR imaging over the
entire GOODS-N field was not available when the target selection
was carried out.    This GOODS-S DRG sample is approximately 
complete for $M_{*} > 10^{11}$ \solm DRG selected galaxies at $z < 3$ 
(Papovich et al. 2006).

Another selection we use to construct our initial massive sample is 
the {\it Spitzer} selected extremely red objects (EROs), otherwise known as Infrared 
EROs (IEROs).  These were first described in Yan et al. (2004), based on 
NIR and {\it Spitzer} data 
within the GOODS fields.  The selection for these objects is 
S$_{\rm \nu}$(3.6 $\mu$m)/S$_{\rm v}(0.9 \mu{\rm m}) > 20$. These objects
were found by Yan et al. (2004), based on SED fits, to have a mixture
of old and younger populations.  Note that selecting galaxies in this
way ensures that they are massive
given their brightness in the IR. However, because they are selected
with {\it Spitzer} imaging, which has a large PSF, resulting in 
potential confusion from neighboring objects, this selection can
have issues with 
contamination
from other galaxies. Hence any galaxies which would satisfy the criteria
but are too close to another galaxy will not be included simply due to
the problem of confusion.

Another method we use to select distant galaxies is through the BzK approach, 
which is described in Daddi et al. (2007) in terms of selection within the 
GOODS fields.    The selection for these objects is slightly more complicated
than that of the DRGs or IEROs, since they are selected through colours using 
the $B$, $z$ and $K$-bands together.  This method proposes to 
separate evolved galaxies or passive pBzKs, and those
which are star forming, or sBzKs.  The selection for these galaxies is
done through the quantity $BzK$, defined using these three bands by:

\begin{equation}
BzK = (z-K)_{\rm AB} - (B-z)_{\rm AB}.
\end{equation}

\noindent Star forming galaxies at $z > 1.4$ are proposed to have 
$BzK > -0.2$. The redder, possibly more evolved galaxies, are found
through the selection $BzK < -0.2$ and $(z-K)_{\rm AB} > 2.5$. For the
BzK sample we use, the selection is somewhat more limited than for
the other colour selections as these sources were selected down
to $K = 20.5$ Vega in the North and $K = 22$ in the South.  We utilise
 photometric redshifts and stellar masses of the galaxies selected
through these methods to identify
and study these colour selected populations taken directly from Papovich
et al. (2006), Yan et al. (2004) and Daddi et al. (2007).  

Our initial massive galaxy sample from which we optimise our 
NICMOS pointings are selected through these three methods, with
a further photometric redshift cut of $1.7 < z < 2.9$, and with
a stellar mass cut of $M_{*} > 10^{11}$ \solm.  
In practice our final pointings were chosen by finding the 
locations within the GOODS
fields where the number of these massive galaxies was maximised within the
NIC3 fields.   In total we imaged 45 pre-selected massive 
$M_{*} > 10^{11}$ \solm
galaxies at $1.7 < z < 2.9$ in the GOODS-N, and 35 in the GOODS-S. 

Galaxies selected in other ways were also used to optimise the 
number of galaxies in each NIC3 pointing, although each pointing was designed
to have at least one massive galaxy with the properties above. 
These `additional' galaxies are 
selected through the Lyman-break drop-out method utilising $B$, $V$ and $i$
drop-outs, the BX/BM selection,
as well as sub-mm galaxies from Greve et al. (2008).  Each NIC3 pointing
contained between four to 19 of each of these galaxy types. 
Figure~1 shows our field layout within the GOODS fields with the different
galaxy types shown as different colours and symbols, and Figure~2 shows
a typical NICMOS NIC3 pointing of one of our fields.

\begin{figure*}
 \vbox to 200mm{
\includegraphics[angle=0, width=120mm]{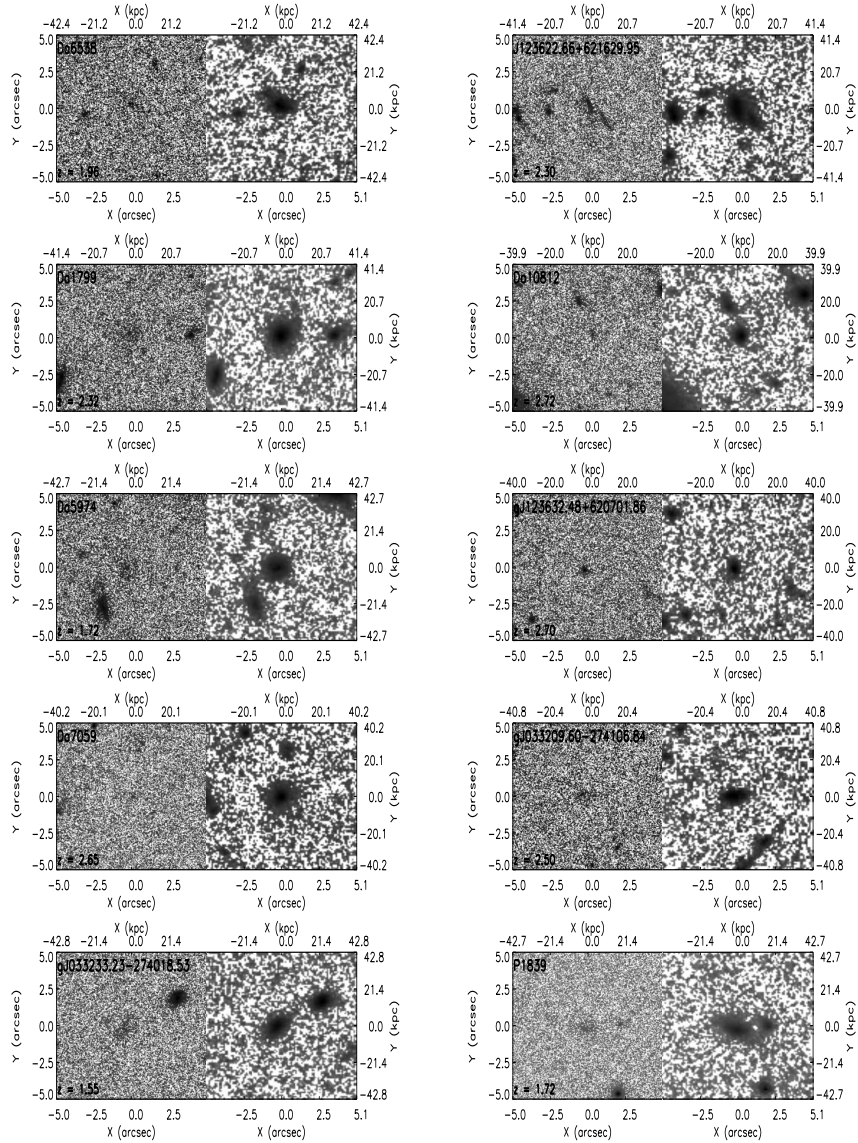}
 \caption{Montage of example GNS massive galaxies with $M_{*} > 10^{11}$ \solm.  
Shown on the left-hand side of each image is the galaxy in the ACS $z$-band, while
the right hand side shows the NICMOS NIC3 $H_{160}-$band view of the same galaxy. 
The sizes in kpc is on the top and in arcsec on the bottom.}
} \label{sample-figure}
\end{figure*}

Tables~1 and 2 list the statistics and positions of our 60 pointings, with
30 NIC3 pointings in the GOODS-North field, and 30 in GOODS-South.  We 
also list 
the number of various other types of galaxies within each of these fields.
Tables~3 and 4 list the initial massive galaxies for which we picked our fields,
along with basic information such as their photometric redshifts,
stellar masses, and information on the optical and $H_{160}-$band magnitudes for
these systems.  NICMOS and ACS images
of ten of these massive galaxies are shown in Figure~3.  This data, including
catalogs of sources, redshifts and stellar masses, as well as the original
reduced NIC3 imaging itself is available at 
{\bf http://www.nottingham.ac.uk/astronomy/gns/}.

\vspace{1cm}
\setcounter{table}{0}
\begin{table*}
\begin{minipage}{170mm}
 \caption{The GOODS-N NICMOS Fields with numbers of galaxies of different types}
 \begin{center}
 \label{tab1}
 \begin{tabular}{@{}ccccccccccc}
  \hline
\hline
ID & RA(J2000) & Dec (J2000) & $V$-drops & $i$-drops & IEROs & BzK & sub-mm & BX/BM & Other & Total \\
\hline
   1 &   12  36 31.8  &  62  06 43.7 & 2 &  1 & 4 &  1 &  1 &  0 &  0 & 9  \\
   2  &  12  36 28.8  &  62  08 07.8 & 1 &  1 & 3 &  1 &  0 &  0 &  0 & 6  \\
   3 &   12  36 14.1  &  62  09 48.5 & 3 &  0 & 5 &  1 &  0 &  0 &  0 & 9  \\
   4  &  12  36 18.5  &  62  09 03.7 & 4 &  0 & 2 &  1 &  0 &  0 &  0  &  7   \\
   5 &   12  37 00.3  &  62  09 09.8 & 0 &  0 & 1 &  1 &  0 &  2 &  0  &  4   \\
   6 &   12  36 41.7  &  62  10 02.3 & 1 &  0 & 3 &  2 &  0 &  1 &  1  &  8   \\
   7 &   12  36 34.3  &  62  14 00.4 & 1 &  0 & 0 &  1 &  1 &  2 &  1  &  6   \\
   8 &   12  38 01.3  &  62  16 15.2 & 1 &  0 & 3 &  2 &  1 &  0 &  0  &  7   \\
   9 &   12  36 54.3  &  62  17 31.9 & 0 &  0 & 1 &  1 &  0 &  2 &  0  &  4   \\
  10 &   12  37 35.9  &  62  20 42.9 & 0 &  0 & 2 &  1 &  0 &  0 &  0  &  3   \\
  11 &   12  37 11.0 &   62  10 51.6 & 3 &  0 & 3 &  1 &  1 &  8 &  0  &  16   \\
  12 &   12  37 03.8 &   62  11 34.8 & 5 &  2 & 2 &  2 &  0 &  4 &  1  &  16   \\
  13 &   12  36 16.3 &   62  15 32.4 & 2 &  1 & 3 &  0 &  1 &  5 &  3  &  15   \\
  14 &   12  37 05.3 &   62  15 00.0 & 5 &  0 & 0 &  1 &  0 &  4 &  2  &  12   \\
  15 &   12  37 13.2 &   62  11 56.4 & 2 &  0 & 3 &  1 &  1 &  3 &  1  &  11   \\
  16 &   12  37 13.9  &  62  15 43.2 & 7 &  0 & 1 &  0 &  0 &  3 &  1  &  12   \\
  17 &   12  36 13.4  &  62  10 37.2 & 4 &  0 & 3 &  0 &  1 &  2 &  0  &  10   \\
  18 &   12  36 50.9  &  62  15 00.0 & 5 &  1 & 2 &  1 &  0 &  1 &  3  &  13   \\
  19 &   12  36 10.1  &  62  08 52.8 & 2 &  0 & 7 &  0 &  1 &  0 &  0  &  10   \\
  20 &   12  36 21.6  &  62  16 19.2 & 2 &  0 & 3 &  0 &  0 &  1 &  1  &  7   \\
  21 &   12  36 20.6  &  62  14 27.6 & 4 &  0 & 3 &  0 &  0 &  2 &  1  &  10   \\
  22 &   12  36 32.2  &  62  09 57.6 & 4 &  0 & 4 &  1 &  0 &  1 &  0  &  10   \\
  23 &   12  36 56.6  &  62  08 16.8 & 5 &  1 & 1 &  0 &  0 &  2 &  0  &  9   \\
  24 &   12  37 09.6  &  62  20 45.6 & 2 &  0 & 3 &  1 &  1 &  1 &  2  &  10   \\
  25 &   12  36 25.0  &  62  10 30.0 & 4 &  0 & 5 &  0 &  0 &  0 &  0  &  9   \\
  26 &   12  36 45.1  &  62  16 04.8 & 2 &  0 & 2 &  0 &  0 &  5 &  0  &  9   \\
  27 &   12  36 50.2  &  62  19 04.8 & 2 &  0 & 3 &  1 &  1 &  0 &  1  &  8   \\
  28  &  12  37 18.2  &  62  12 32.4 & 0 &  0 & 1 &  1 &  0 &  5 &  1  &  8   \\
  29 &   12  37 49.7  &  62  14 16.8 & 1 &  0 & 3 &  2 &  1 &  0 &  0  &  7   \\
  30 &   12  35 57.6  &  62  10 39.4 & 0 &  1 & 2 &  0 &  0 &  0 &  1  &  4   \\
\hline
\end{tabular} \\
\end{center} 
Notes. Listed are the centre positions of each NIC3 pointing for the GNS in terms
of RA and Dec. Also
listed are the number of $V-$drops, $i-$drops, IEROs, BzKs and DRGs.  Note that the colour selected types listed here are not just galaxies with $M_{*} > 10^{11}$ \solm, but all galaxies that meet the criteria outlined in Section 2.  We also
include an `Other' column which includes the total summation of: sub-mm galaxies,  
galaxies with {\it Spitzer} infrared spectrograph coverage, and BM/BX galaxies from Reddy et al. (2008).  
The total number of galaxies we used to pick these fields is also
shown.
\end{minipage}
\end{table*}

\vspace{1cm}
\setcounter{table}{1}
\begin{table*}
\begin{minipage}{170mm}
\begin{center}
 \caption{The GOODS-S NICMOS Fields with numbers of galaxies of different types}
 \label{tab1}
 \begin{tabular}{@{}cccccccccc}
  \hline
\hline
ID & RA (J2000) & Dec (J2000) & $V$-drops & $i$-drops & IEROs & BzK & DRGs & Other & Total \\
\hline
   1  &   03  32  23.5  &  $-27$  48  18.0  &   3 &  2 & 5 &  1 &  5  &   0  & 16 \\
   2  &   03  32  24.2  &  $-27$  43  04.8  &   2 &  1 & 8 &  0 &  2  &   3  & 16 \\
   3  &   03  32  22.8  &  $-27$  45  46.8  &   3 &  2 & 4 &  1 &  3  &   0  & 13    \\
   4  &   03  32  30.0  &  $-27$  48  18.0  &   2 &  1 & 6 &  1 &  2  &   1  & 13    \\
   5  &   03  32  42.2  &  $-27$  49  22.8  &   4 &  1 & 2 &  1 &  0  &   2  & 10     \\
   6  &   03  32  54.2  &  $-27$  51  10.8  &   2 &  0 & 8 &  1 &  2  &   0  & 13    \\
   7  &   03  32  15.6  &  $-27$  41  38.4  &   3 &  1 & 4 &  0 &  2  &   1  & 11    \\
   8  &   03  32  21.8  &  $-27$  42  25.2 &    1 &  0 & 3 &  1 &  1  &   0  & 6    \\
   9  &   03  32  29.3  &  $-27$  53  20.4  &   2 &  0 & 5 &  0 &  3  &   0  & 10    \\
  10  &   03  32  30.7  &  $-27$  54  14.4  &   1 &  0 & 3 &  2 &  4  &   0  & 10    \\
  11 &    03  32  45.8  &  $-27$  53  31.2   &  2 &  0 & 4 &  0 &  3  &   0  & 9    \\
  12 &    03  32  53.8  &  $-27$  52  26.4  &   2 &  0 & 3 &  1 &  3  &   0  & 9    \\
  13 &    03  32  12.0  &  $-27$  43  04.8  &   1 &  0 & 4 &  0 &  3  &   0  & 8    \\
  14  &   03  32  14.9 &   $-27$  52  26.4  &   3 &  0 & 3 &  1 &  1  &   0  & 8    \\
  15 &    03  32  16.3  &  $-27$  50  38.4  &   3 &  1 & 3 &  1 &  1  &   0  & 9    \\
  16  &   03  32  17.8  &  $-27$  51  32.4  &   1 &  0 & 3 &  1 &  3  &   0  & 8    \\
  17   &  03  32  32.9  &  $-27$  40  22.8   &  3 &  2 & 2 &  0 &  0  &   1  & 8    \\
  18  &   03  32  07.7  &  $-27$  41  38.4  &   3 &  0 & 3 &  1 &  0  &   0  & 7    \\
  19  &   03  32  09.8  &  $-27$  48  18.l  &   4 &  1 & 2 &  1 &  0 &    0  & 8    \\
  20  &   03  32  11.3  &  $-27$  41  16.8  &   2 &  0 & 6 &  0 &  0 &    0  & 8    \\
  21  &   03  32  14.9  &  $-27$  49  49.1  &   3 &  1 & 2 &  0 &  1 &    0  & 7    \\
  22  &   03  32  28.6  &  $-27$  54  57.6  &   2 &  1 & 2 &  1 &  1 &    0  & 7    \\
  23  &   03  32  28.6  &  $-27$  52  15.6   &  1 &  2 & 2 &  1 &  2  &   0  & 8    \\
  24  &   03  32  31.4  &  $-27$  50  16.8   &  2 &  1 & 3 &  1 &  0  &   0  & 7    \\
  25  &   03  32  24.2  &  $-27$  55  34.7   &  3 &  1 & 2 &  0 &  1  &   2  & 9    \\
  26 &    03  32  43.7  &  $-27$  42  46.8  &   3 &  0 & 3 &  1 &  0  &   0  & 7    \\
  27 &    03  32  32.2 &   $-27$  55  01.2  &   1 &  0 & 1 &  2 &  0  &   0  & 4    \\
  28 &    03  32  07.7  &  $-27$  46  40.8  &   2 &  3 & 2 &  0 &  0  &   0  & 7    \\
  29  &   03  32  33.0  &  $-27$  45  43.9  &   2 &  0 & 1 &  1 &  1  &   0  & 5    \\
  30  &   03  32  39.3  &  $-27$  42  48.3  &   4 &  2 & 0 &  1 &  1 &    0  & 8    \\
\hline
\end{tabular} \\ 
\end{center} 
Notes. Listed are the centre positions of each NIC3 pointing for the GNS in terms
of RA and Dec. Also
listed are the number of $V-$drops, $i-$drops, IEROs, BzKs and DRGs.   Note that the colour selected types listed here are not just galaxies with $M_{*} > 10^{11}$ \solm, but all galaxies that meet the criteria outlined in Section 2. 
We also include an `Other' column which includes the total summation of: 
sub-mm galaxies, galaxies with {\it Spitzer} infrared spectrograph coverage, and BM/BX galaxies from Reddy et al. (2008).  The total number of galaxies we used to pick these fields is also
shown.
\end{minipage}
\end{table*}

\subsection{Observational Parameters and Data Reduction}

For our observations, we used the NIC3 camera in the $H$(F160W) ($H_{160}$) 
band, with a depth of
3 orbits per pointing.  With this exposure time we predicted that we 
would reach $H_{160}$ = 26.5 (AB) at 5 $\sigma$  for an extended source 
within a 0.7 arcsec diameter.   This imaging, combined with ground-based
data at a similar depth, is optimal for measuring photometric redshifts 
at $z \sim 2$,
where the Balmer break occurs for galaxies at these redshifts, 
and for finding $z-$band drop-outs ($z > 6$) as candidate high$-z$
galaxies (Bouwens et al. 2010).    

Our data were processed with the NICMOS reduction package \texttt{NICRED.py}
v1.0.  A
detailed description of this package can be found in Magee, Bouwens, \&
Illingworth (2007).
\texttt{NICRED} v1.0 handles all pipeline processing steps currently
recommended by STScI for NICMOS data. 
Basic calibration, including zero read correction, bad pixel masking,
noise calculation, dark current subtraction, linearity correction,
flat field correction, photometric calibration, and cosmic ray
identification, was handled with the IRAF task \texttt{CALNICA}.  Pedestal 
removal and bias subtraction was performed with the 
IRAF task \texttt{PEDSKY}.  

The NICMOS data was taken with three dithers per
orbit using a point spacing of 5.06, for a total of nine dithers, which 
was then used to drizzle the data into the final product.  Exposure times 
are roughly 8100 seconds per target in the F160W band.
Each exposure was cleaned of any South Atlantic Anomaly
signatures using the algorithm of
Bergeron \& Dickinson (2003, ISR), as implemented in the STScI-Python task
\texttt{saaclean.py}.   Correction for the ``Mr. Staypuft'' anomaly was also
made using a python task equivalent to STScI-python 
task \texttt{puftcorr.py}.  

To improve the overall flatness of individual frames, we
median-stacked all of the frames associated with the program
(after masking out individual sources) to create a super median frame.
We then subtracted this median from each of the individual exposures.
Each exposure was also corrected for the NICMOS count rate non-linearity
as identified by Bohlin et al. (2005) and later more thoroughly
characterized by de Jong et al. (2006).  This latter correction was
made using the python task \texttt{nonlincor.py}.  

Because our NICMOS images were
taken in a single visit, we did not attempt to correct the relative
astrometry of the individual frames to improve the overall alignment
solution, although we did adjust the astrometry by 0.3\arcsec\, in declination
when aligning with the GOODS ACS v2.0 imaging.  
In preparation for the final image combination process,
inverse weight maps were computed for each exposure based upon their
individual exposure times, the reference darks, and flat fields.
Finally, the individual exposures were combined into a final rectified
frame with multidrizzle, rejecting any pixel in an exposure that was
more than 4 $\sigma$ away from the median defined by the stack.  We used
a threshold of 3.5 $\sigma$ for the rejection threshold for pixels adjacent 
to an already rejected pixel.

We calculated that zero-points for the $H_{160}$ images are 25.17 AB mag, and 
include a correction for the NICMOS non-linearity count 
rate\footnote{This zero point
differs from that given in the headers of the NICMOS data themselves,
because the imaging data are expressed in electrons s$^{-1}$ rather DN 
$s^{-1}$, with a gain of 6.5.}. In total we obtained 60 pointings for our 
observations, each one
roughly centred on a massive galaxy at $z = 1.7 - 2.9$.  The field of view of
each of these images is 51.2 arcsec on a side with a subsampled pixel scale
for the final drizzled mosaics of
0.1 arcsec pixel$^{-1}$. We measure that the FWHM for our images is roughly
$0.3$ arcsec.

The depth of our data was determined by placing random apertures
throughout the images and determining what fraction of simulated sources 
can be retrieved,
and by measuring the RMS noise at various positions in
the imaging.  Using this method we find that the depth at the 
5 $\sigma$ limit is 
26.8 AB mag using a 0.7 arcsec-diameter aperture.  This is similar to our 
initial estimate based on the depth of our imaging.

\subsection{SExtractor Detections and Photometry}

After our images were reduced to their three orbit depths we then carried
out image detection and photometry with the SExtractor package.  This is
now a standard method for detecting galaxies within imaging, and we only
give a brief overview of the methods which we used.  The photometry
was also done within SExtractor, both for total magnitudes as
well as magnitudes measured with a series of apertures.

\begin{figure}
 \vbox to 120mm{
\includegraphics[angle=0, width=90mm]{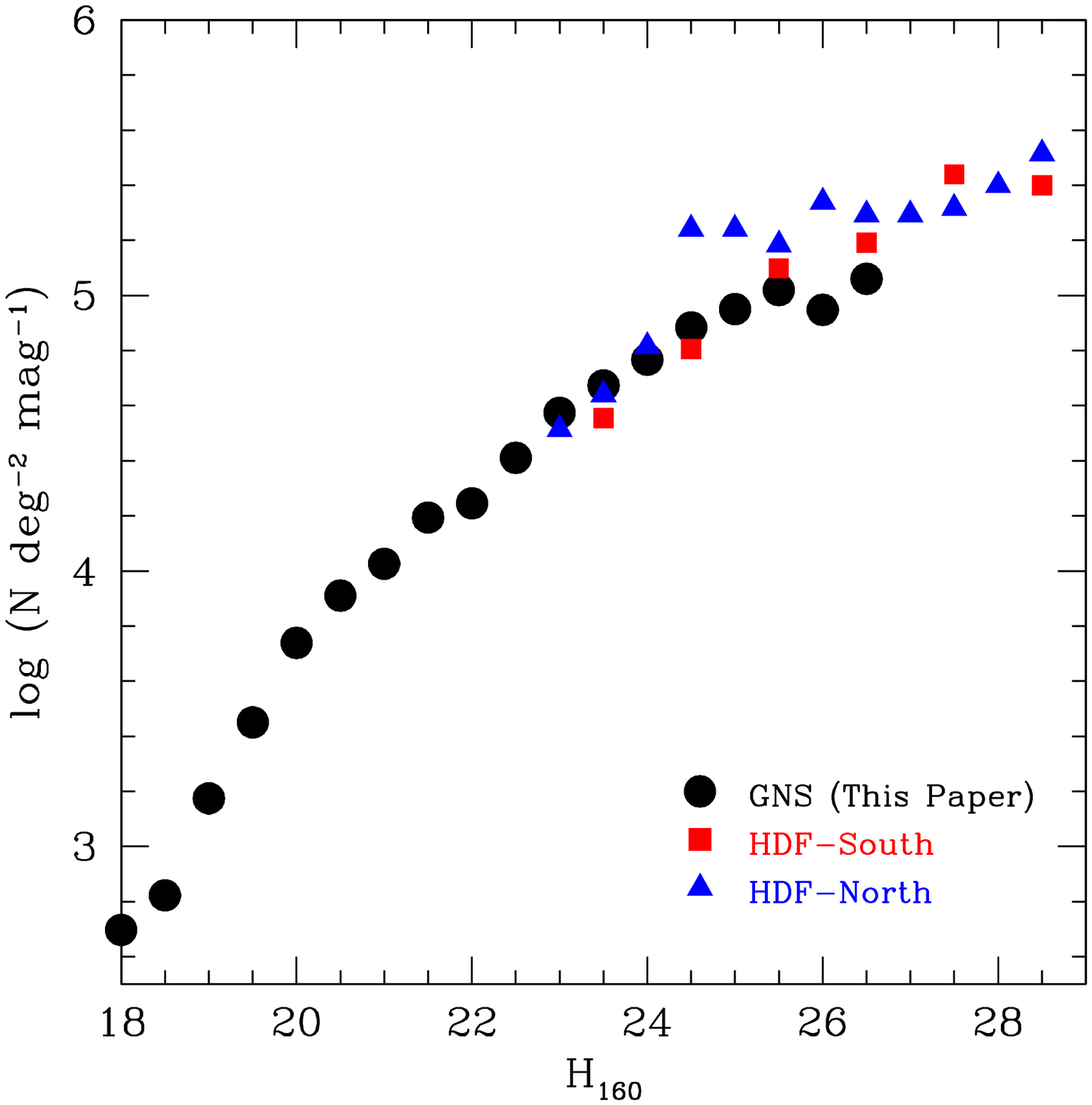}
 \caption{Number counts within our GNS $H_{160}$ band imaging.  Shown
for comparison are number counts from the HDF-North and HDF-South
imaging which were taken using the same camera and filter.  Data for
the {\it Hubble} Deep Field (HDF) South originates from Metcalfe et al.
(2006) and the the HDF-N number counts are from Thompson et al. (1999).}
} \label{sample-figure}
\end{figure}

The basic idea behind SExtractor is to detect objects within an astronomical
image and to carry out basic analyses of the photometry and shapes of
these objects, typically galaxies and stars. Using SExtractor on an interactive basis we were able to find an optimal detection and deblending method that accounted for nearly all objects that can be identified as separate galaxies and stars by eye.  Weight maps of the exposure times were used within the detection
procedure.
Overall we find a total of 8298 galaxies within our SExtracting and
cleaning process.

We constructed, an optically matched catalogue of ACS imaging in the 
$BViz$ bands
based on the positions of the galaxies in the NICMOS catalog. Photometry 
in $B$, $V$, $i$ and $z$ bands are available for sources down to a 5 
$\sigma$ limiting 
AB magnitude of $z \sim 27.5$ from the original ACS GOODS survey (Giavalisco
et al. 2004) using v2.0 data products.  We used
the positions of objects in our NICMOS catalogue to match with $BViz$
photometry from the ACS v2.0 data.  We used {\em AUTOMAG} magnitudes
to measure the magnitudes for both the $H-$band and the $BViz$ magnitudes, 
which accounts for the total amount of light within each galaxy at every
wavelength.   We used these magnitudes to obtain $BVizH$ spectral energy
distributions for every source.  These SEDs are used for colour measurements
as well as for SED fitting for photometric redshifts and stellar mass
calculations.  A large fraction of our sources, 1219 out of 8298, have no
counterpart in the $z-$band down to our limit of $z_{850} = 27.5$.

  After accounting for a well-known 0.3\arcsec offset in the
declination direction between the NICMOS and ACS v2.0 data, we find that the
average offset between the ACS positions and the NICMOS sources they
are identified with is 0.06$\pm$0.04\arcsec.  
We then later use this optical
and NIR matched catalog to derive properties such as photometric redshifts
as well as stellar masses for each of our galaxies.  This also allows for
us to search for drop-out galaxies which may be at ultra-high redshift
(Bouwens et al. 2010).  

We decided to only use {\it HST} imaging for our photometric catalogue when
calculating redshifts and stellar masses to ensure a high fidelity in our 
photometry. 
While the GOODS fields have imaging at many ground-based wavelengths 
(Giavalisco et al. 2004), this imaging is often at a similar wavelength range
to an ACS+NICMOS catalog, with the exception of a few
pass-bands, such as the $U$-band and $K$-band.   The accuracy of our matched
photometry is very high, and our depth much greater than this ground
based imaging, and thus to obtain a cleaner measurement we have limited our
analysis to these five bands. Furthermore, we do not use {\it Spitzer} IRAC
photometry for our galaxies (although we have matched these) simply because
of issues due to contamination and deblending which can be substantial for
galaxies which are separated by less than the PSF of the IRAC imaging 
(several arcsec in FWHM), 
making measurements of our photometry and stellar masses much more difficult.

\subsubsection{Number Counts}

In Figure~4 we present the number counts for our $H_{160}$-band imaging, with
a comparison to counts from the {\it Hubble} Deep Field South (Metcalfe et al.
2006) and the HDF-N (Thompson et al. 1999).  Based on a comparison to 
these $H_{160}$ counts, we are
roughly complete in our galaxy selection to roughly 
$H_{160}$ = 25.5.  The scatter in the counts
at the faint end are most certainly due to cosmic variance
effects, given the small field of view of these previous surveys.
There are, however, some
differences, particularly at the faint end of the counts, which can
also be seen by comparing the HDF-S and HDF-N counts.  This 
shows that we are obtaining similar photometric quality to these previous
deeper NICMOS pointings and our reduction and detection processes are
consistent with previous work.

\section{Derived Parameters}

There are two primary catalogues used within the GNS.  The first is the
initial catalogue of massive galaxies, selected by the methods
described in \S 2.2.  The other is the SExtractor based catalog of the 
survey
based on the $H_{160}$-band imaging. This catalogue is an $H_{160}$-band
catalogue of every object which is imaged
within the NIC3 survey, regardless of mass and brightness. 
These two catalogues will be used for different
purposes throughout this study, and in the follow-up papers, with
detailed analyses of various aspects of this work. We
describe in this section the redshift and stellar mass data we use to 
construct the first sample from which our initial targets were chosen.
We also describe in detail the redshifts and stellar masses
derived from the new optical+NIR catalog which we constructed using
our $H_{160}$-band selected objects matched to the optical ACS
photometry.

\subsection{Photometric Redshift Measurements}

To obtain photometric redshifts, the NICMOS $H_{160}$ band sources are
matched to the catalogue of optical sources in the GOODS-ACS fields as
described in \S 2.4.
Photometric redshifts are then obtained by fitting template spectra to the 
$BVIzH$ photometric data points.  We do not include other wave-bands or
ground-based data; we omit these so that we
can obtain the highest fidelity photometry not affected by zero-point
random and systematic errors, background noise, or confusion with other 
sources, as described in \S2.3.  
The degeneracy in color-redshift space is problematic, especially 
when few filters are available. To cope with this effect 
we used two different approaches: the standard $\chi ^2$ minimisation 
procedure, using {\em HYPERZ} (Bolzonella et al. 2000),
and a Bayesian approach using the {\em BPZ} method (Benitez 2000).

\begin{figure*}
 \vbox to 150mm{
\includegraphics[angle=-90, width=180mm]{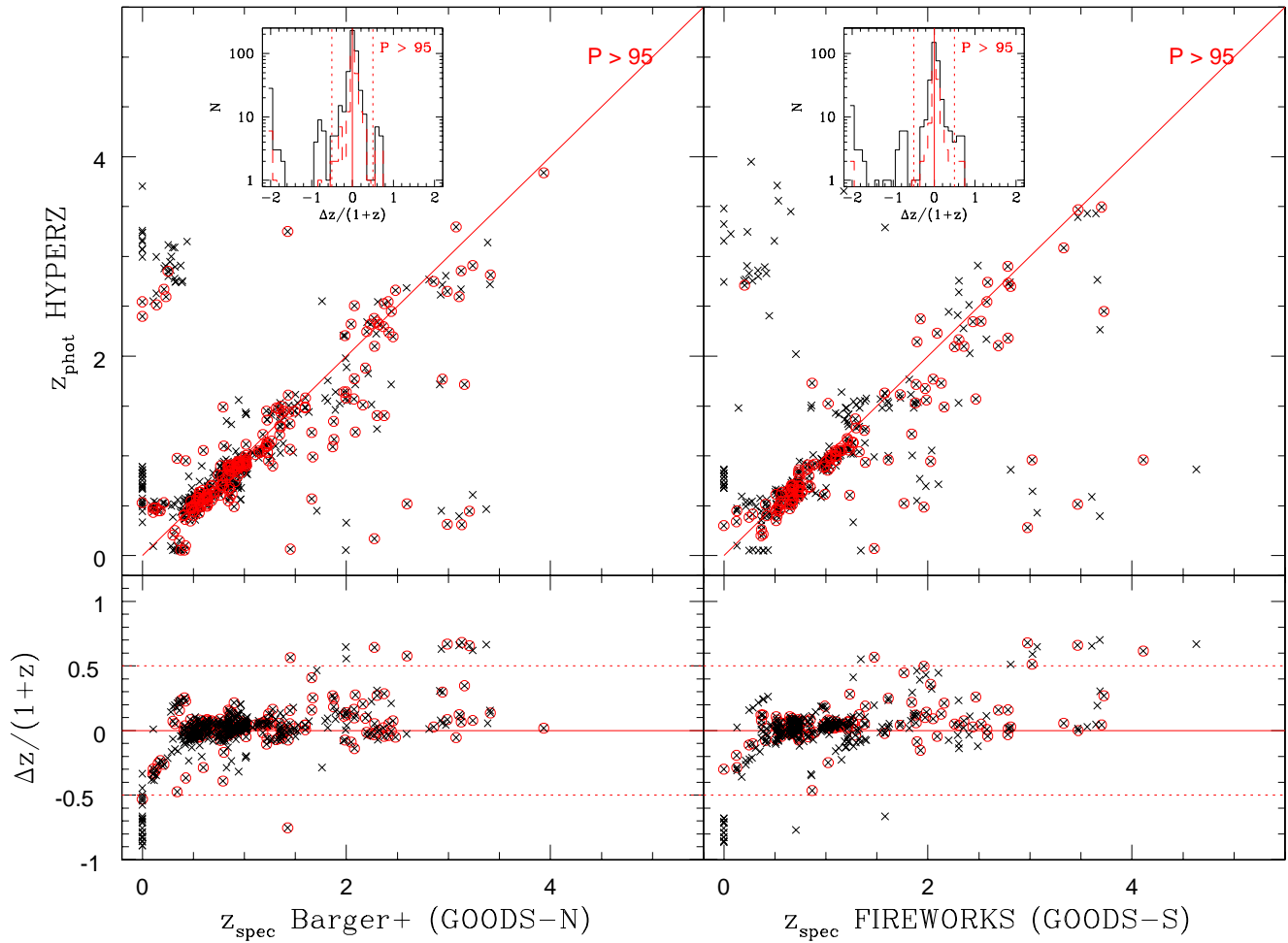}
 \caption{Reliability of photometric redshifts: {\it Top panels:} photometric vs. spectroscopic redshifts in the GOODS-N (left) and GOODS-S (right) fields for galaxies with redshift probabilities ($P$) greater than 95 percent. The insets show the distribution of $\Delta z/(1+z)$ for all photometric redshifts (black) and high probability redshifts only (red, long dashed). {\it Bottom panels:} $\Delta z/(1+z)$ dependence on redshift. Black symbols show all redshifts, red symbols high probability redshifts only. The dashed lines in all the panels and subpanels, show a limit for catastrophic outliers at $|\Delta z/(1+z)| > 0.5$.}
\vspace{5cm}
} \label{sample-figure}
\end{figure*}

The synthetic spectra used by {\em HYPERZ} are constructed with the Bruzual \& 
Charlot (2003; BC03) evolutionary code, representing roughly the different  
types of galaxies found in the local Universe. We use five template spectra 
corresponding to the spectral types of:  E, Sa, Sc and Im, as well as a single 
star burst model. The reddening law is taken from Calzetti et al. (2000).
The code then computes the most likely redshift solution in the parameter 
space of age, metallicity, and reddening. The best fit redshift and 
corresponding probability are then output together with the best fit 
parameters of spectral type, age, metallicity, $A_V$ and secondary solutions
of these. 

The Bayesian approach of Benitez (2000) uses a similar template fitting 
method, as well as using an empirical rather than synthetic template SEDs. 
The main difference from {\em HYPERZ} is that it does not rely on the 
maximum likelihood 
of the redshift solution in the parameter space as described above. 
Instead it uses additional empirical information about the likelihood of a 
certain combination of parameters, also known as prior information, or
priors. The redshift solution with the maximum likelihood is determined 
after weighting the probability of each solution by the additional 
probability determined 
from the prior information. In our case the prior is the distribution of 
magnitudes for different morphological types as a function of redshift, 
obtained from {\it Hubble} Deep Field North (HDF-N) data (Benitez 2000). In 
short, the code not only determines the best fit redshift and spectral 
type, but also takes into account how likely is it to find a galaxy of that 
spectral type and magnitude at the given redshift. 

\subsection{Comparison with Spectroscopic Redshifts}

Since the spectroscopic redshifts of sources in the two GOODS fields 
(North and South) were taken from different compilations of data, we compare 
them separately to the photometric redshifts from their respective fields
in this section.

Spectroscopic redshifts of sources in the GOODS-N field were compiled by 
Barger et al. (2008), whereas in the GOODS-S field spectroscopic 
redshifts are taken from the FIREWORKS compilation (Wuyts et al. 2008). We 
matched these catalogues to our photometric catalogue, 
obtaining 537 spectroscopic redshifts for our sources 
in GOODS-N and 369 in GOODS-S. The mean separation between photometric and 
spectroscopic sources is $0.41$ $\pm 0.06$ arcsec in the
 GOODS-N field and $0.13$ $\pm 0.05$ arcsec in the GOODS-S field. 

The reliability of photometric redshift measures is often defined by 
$\Delta z/(1+z) \equiv (z_{\rm spec}- z_{\rm phot})/(1+z_{\rm spec})$. 
In the following 
we compare the median error ($\langle \Delta z/(1+z) \rangle$) and rms 
scatter ($\sigma$) as well as the fraction of catastrophic outliers, 
i.e., galaxies with both $|\Delta z/(1+z)| > 0.5$ and $> 0.2$, obtained by 
the two methods described above. 

We find good agreement between photometric and spectroscopic redshifts 
for both codes.  However, {\em HYPERZ} gives slightly better results, although 
{\em BPZ} gives more high probability ($P$) redshifts.
Using {\em HYPERZ}, we obtain the following results: sources in the GOODS-N 
field have an $ \Delta z/(1+z) \rangle = 0.027$, with a scatter of 
$\sigma = 0.04$ (222 out of 537 galaxies with $P > 95$ percent). Sources 
in the GOODS-S field show similar values: 
$\langle \Delta z/(1+z) \rangle = 0.043$ and $\sigma = 0.04$ (134 of 
369 with $P > 95$ percent). {\em BPZ} gives slightly higher errors and scatter: $\langle 
\Delta z/(1+z) \rangle$ = 0.07 and $\sigma = 0.05$ for GOODS-N (475 galaxies) 
and $\langle \Delta z/(1+z) \rangle = 0.07$ and $\sigma = 0.06$ for GOODS-S 
(317 galaxies).  We find that galaxies with lower probability redshifts
give similar accuracy when compared to spec-zs.  We therefore use 
all  of the photometric
redshifts calculated within our analysis.  
The fraction of catastrophic outliers 
is $\sim 6$ percent for both codes. It rises to $\sim 16$ percent for galaxies 
with $|\Delta z/(1+z)| > 0.2$.  Surveys of high$-z$ galaxies using 
multiple medium band NIR filters find 
photometric redshifts similar to ours with 
$\Delta z/(1+z) \sim 0.2$ (van Dokkum et al. 2009).

The relatively good agreement between photometric and spectroscopic redshifts 
is shown in Figure~5. The photometric redshifts of the {\em HYPERZ} 
code are plotted against the spectroscopic redshifts of GOODS-N and GOODS-S. 
Objects with a high probability value of $z_{\rm phot}$ are encircled in red. 
Most outliers, especially  at low redshift do not have a high probability. 
The lower panel shows the $\Delta z/(1+z)$ dependence on redshift 
$z_{\rm spec}$, 
where there is no clear trend, or bias, with redshift, with the possible 
exception of a slight trend to underestimate redshifts at high$-z$.

We are also interested in how good our photo$-z$s are with respect to redshift,
as well as within our selection method for our sample, which uses the 
$H_{160}$-band. Thus Figure~6 shows the dependence of $\Delta z/(1+z)$ 
on $H_{160}$-band magnitude. {\em HYPERZ} and {\em BPZ} results are plotted
in red 
and blue, respectively.   Only high probability redshift are used in 
this figure. The median error and rms 
scatter are computed in each magnitude are shown.   The figure shows the 
slightly better performance of {\em HYPERZ}, which is also visible in the 
fraction of outliers with $|\Delta z/(1+z)| > 0.5$. The redshift error is 
stable up to faint magnitudes of $H_{\rm 160} \sim 24$, as is the fraction of 
outliers.  {\em HYPERZ} is likely giving a superior result over 
{\em BPZ} due to the limited
redshifts we can use as the Bayesian training set, which thus limits the 
reliability of redshifts and/or types of galaxies for which no spec$-z$s are 
available.  See also Conselice et al. (2007a) for a more general discussion 
about using different types of photo$-z$s for galaxies with different 
properties.

The comparison of our results with photometric redshifts already available 
for the brighter part of our sample shows that our photometric redshifts are 
of comparable quality to those using many more photometric bands, although
these necessarily are from ground-based and/or {\it Spitzer} imaging. 
Photometric redshifts taken from the FIREWORKS 
compilation have a median difference of $\Delta z/(1+z) = 0.037$ and an rms 
scatter of $\sigma = 0.028$ in comparison to our photometric
redshifts.  While we find good agreement between our photometric redshifts
and previously published spectroscopic redshifts, it must be noted that
most of these galaxies are fairly bright and it remains to be determined
whether our agreement would be as good for much fainter galaxies.

\begin{figure*}
 \vbox to 140mm{
\includegraphics[angle=0, width=120mm]{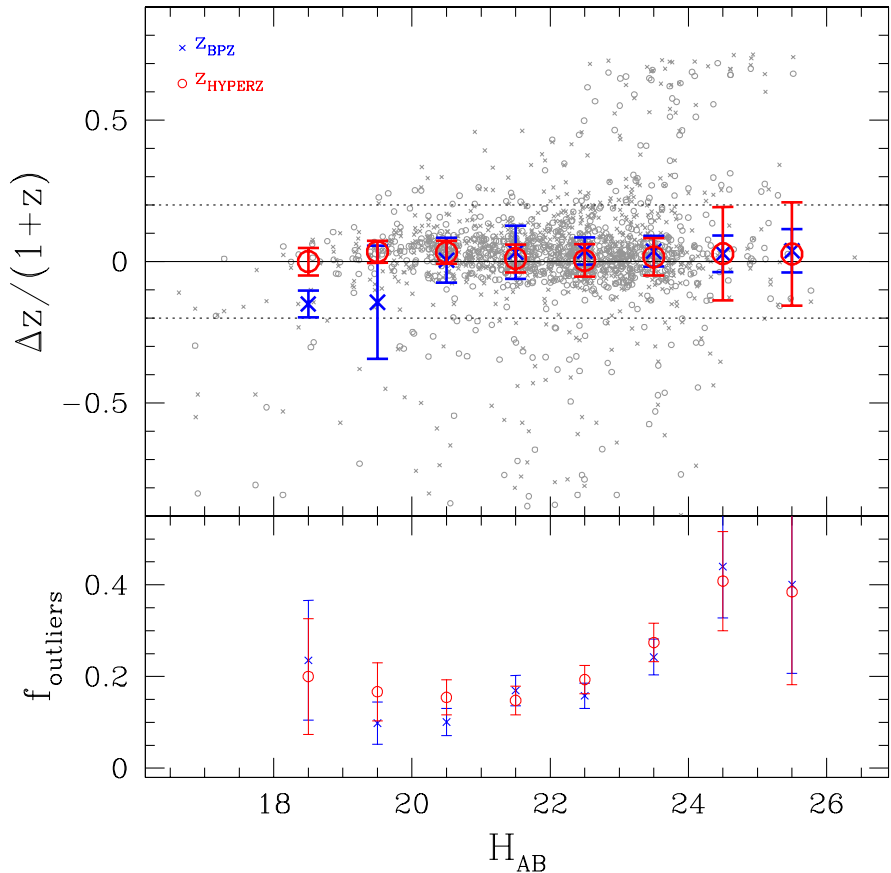}
 \caption{Dependence of $\Delta z/(1+z)$ on $H_{160}$ magnitude.{\it Top panel:} $\Delta z/(1+z)$ vs. $H_{160}$-band magnitude for {\em HYPERZ} (red circles) and {\em BPZ} (blue crosses) results. Median values in each magnitude bin (width = 1 mag) are plotted as solid lines, with the rms scatter shaded in the respective color. {\it Bottom panel:} fraction of catastrophic outliers $|\Delta z/(1+z)| > 0.5$ as a function of $H_{160}$-band 
magnitude. Only high probability redshifts are used in this plot.}
} \label{sample-figure}
\end{figure*}

\subsection{Stellar Masses}

We calculate stellar masses for our galaxies within our global 
$H_{160}$-selected sample through 
the use of our optical+NIR photometry, using our own stellar mass code.  
The method we use to measure stellar masses involves fitting the
photometric points, based on a given redshift, to simulated magnitudes based 
on different star formation histories, and constructing a distribution of
likely stellar masses, as well as other parameters such as
rest-frame optical colours, ages of the stellar population, metallicity,
dust extinction, and so on.  While these non-stellar mass parameters are
degenerate, the stellar mass in these calculations is robust (Bundy
et al. 2006).

In detail, the basic stellar mass fitting method consists of fitting a grid of 
model SEDs constructed
from BC03 stellar population synthesis models, with
different star formation histories. We use an exponentially declining model
to characterise the star formation history, with various ages, 
metallicities, and dust contents used for different models.  These models are 
parameterised by an age, and an e-folding time for parameterising the 
star formation history ($\tau$), and star formation rate ($\psi$) such that

$$ {\rm \psi (t)} \sim\, {\rm \psi_{0}} \times {\rm \exp(-t/\tau)}.$$

\noindent The values of $\tau$ are uniformly selected from a range between 
0.01 and 10 Gyr, while the age
of the onset of star formation ranges from 0 to 10 Gyr. The metallicity
ranges from 0.0001 to 0.05 (BC03), and the dust content is parametrised
by $\tau_{\rm v}$, the effective $V$-band optical depth, for which we 
use values
$\tau_{\rm v} = 0.0, 0.5, 1,$ and 2.     Although we vary several parameters,
the resulting stellar masses from our fits do not depend strongly on the
various selection criteria used to characterise the age and the metallicity
of the stellar population (e.g., Papovich et al.
2006; Bundy et al. 2006, 2008; Conselice et al. 2007a).

It is important to realise that these  parameterisations are
fairly simple, and it remains possible that stellar mass from
older stars is missed under brighter, younger, populations or from
an incorrect star formation parametrisation. For example, 
Papovich et al. (2010) find
that galaxies are increasing in their star formation rate at $z > 2$, although
Papovich et al. (2010) find that this increase measured masses by a factor
of 1.6 at most.   Furthermore, while the majority of our systems are red 
galaxies
it is possible that up to a factor of two in stellar mass is missed in any 
star-bursting blue systems.  However, stellar masses measured through
our technique are roughly the expected factor of five to ten times smaller than
dynamical masses at $z \sim 1$, using a sample of disk galaxies
(Conselice et al. 2005), demonstrating their inherent reliability to within
a factor of two, similar to the estimated errors based on fitting (Bundy
et al. 2006).  Our
method is also the same as that used to trace the evolution of
massive galaxies at lower redshifts $z < 2$ (e.g., Conselice
et al. 2007; Bundy et al. 2006).  Our masses also agree with results
from multiple methods of measuring stellar masses for the same galaxies 
(Papovich et al. 2006; Yan et al. 2004; Daddi et al. 2007).

\begin{figure}
 \vbox to 120mm{
\includegraphics[angle=0, width=90mm]{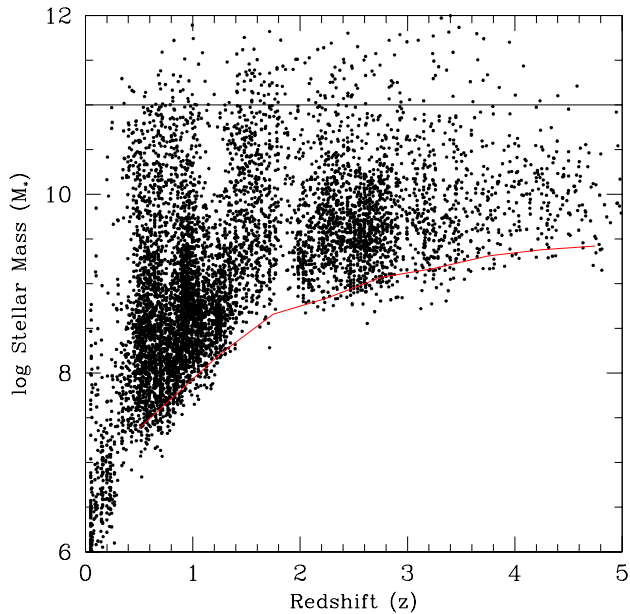}
 \caption{The distribution of redshifts and stellar masses for our
$H_{160}$-selected GNS galaxies. The red solid line shows the evolution of the
minimum stellar mass we could detect at our $H_{160}$-band depth
as a function of redshift for a maximally old stellar population.  
The solid horizontal line shows the log $M_{*} > 11$ limit
in which the primary sample for selection of the GNS galaxies was
carried out.}
} \label{sample-figure}
\end{figure}

We fit the magnitudes derived from these model 
star formation histories to the actual data, to obtain a measurement
of stellar masses using a Bayesian approach.  We calculate 
the likely stellar mass, age, and absolute magnitudes for each galaxy at all 
star formation histories, and determine stellar masses based on this
distribution.  Distributions with larger ranges of stellar masses
have larger resulting uncertainties.  
Typical errors for our stellar masses are 0.2 dex from the width
of the probability distributions.  There are also uncertainties from
the choice of the IMF.  Our stellar masses utilise the
Salpeter IMF. There are additional
random uncertainties due to photometric errors.  The resulting
stellar masses thus have a total random error of $0.2-0.3$ dex,
roughly a factor of two.  

There is also a question as to whether or not our stellar masses
are overestimated because of using the Bruzual \& Charlot (2003)
models.  It has been argued by Maraston (2005), among
others, that a refined treatment of thermal-pulsating asymptotic 
giant (TP-AGB) stars 
in the BC03 models results in stellar masses that can be
too high by a factor of a few. While we consider an uncertainty
of a factor of two in our stellar masses, it is worth investigating
whether or not our sample is in the regime where the effects of
a different treatment of AGB stars, as in e.g., Maraston (2005), will 
influence our mass measurements.  This has been investigated in 
Maraston (2005) who have concluded that galaxy stellar masses computed
with an improved treatment of AGB stars are roughly 
$50-60$ percent lower.

However, the effect
of TP-AGB stars is less important at the rest-frame wavelengths we probe 
than at longer wavelengths, especially in the rest-frame IR. Since the GNS
is $H_{160}$-band selected, and the observed $H_{160}-$band is used as the 
flux in which stellar masses are computed, then the
rest-frame wavelength probed is roughly $\sim 
0.5\, \mu$m at $z \sim 2$. At this wavelength, the effects of TP-AGB stars are 
minimised, as has have shown in previous work using the same type of data, 
and the same code (Conselice et al. 2007). To test this on our
galaxy sample, we utilised
the newer Bruzual and Charlot (2010, in prep) models, which include an improved 
TP-AGB star prescription. From this we
find on average that stellar masses are smaller on average by 
$<$ 0.07 dex  using the newer
models.  At most, the influence of TP-AGB stars will decrease our
stellar masses by 20 percent.  The effect of this would reduce
the number of galaxies within our sample, particularly those
close to the $M_{*} =$ \mass boundary.  This systematic error is however
much smaller than both the stellar mass error we assume (0.3 dex), and
the cosmic variance uncertainties (e.g., Conselice et al. 2007), 
and thus we conclude that it is not
a significant factor within our analysis.  The stellar mass vs. redshift
relation for our sample is shown in Figure~7.  We analyse the stellar mass
function in detail in Mortlock et al. (2010), including how photometric
redshift and stellar mass uncertainties affect the stellar mass function 
up to $z \sim 3.5$, although we give an initial analysis of the
number densities for massive galaxies with M$_{*} > 10^{11}$ \solm later
in this paper (\S 4.3.3).

\begin{figure*}
 \vbox to 110mm{
\includegraphics[angle=0, width=180mm]{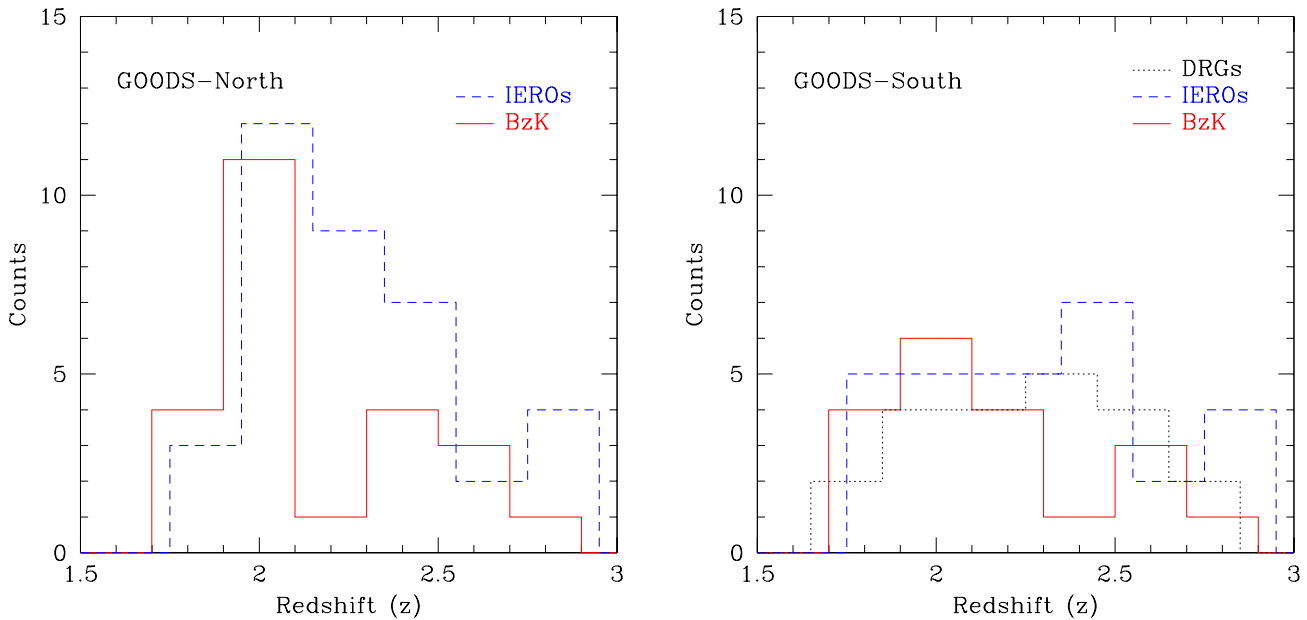}
 \caption{The redshift distribution of galaxies of
different types.  Shown in the left panel are the redshift
distributions for the IEROs and BzKs denoted by dashed
and solid lines, respectively, for the GOODS-N
Field.  The right panel shows a similar trend, except it also
includes the DRG-selected galaxies.  Note that if an
object is selected by multiple methods, that galaxy is plotted
for each type to which it belongs. It appears that the
IEROs are selected from a slighter higher redshift population than
the BzK galaxies.}
} \label{sample-figure}
\end{figure*}

\section{Analysis}

\subsection{Massive Galaxy Selection}

The selection of massive galaxies at high redshift is an important
process which remains difficult due to the inability to easily
acquire spectroscopic redshifts for a sizeable population of 
galaxies at $z > 2$.  At lower redshifts ($z < 1.4$), it is fairly
straightforward to obtain redshifts through spectroscopic surveys
such as DEEP2 or VVDS, combined with deep NIR imaging to measure
stellar masses (e.g., Conselice et al. 2007; Bundy et al. 2006).

As described in our selection method for the GNS fields, there are
a few approaches for determining the massive galaxy population at high
redshifts. These methods typically use a colour selection of some
form, ranging over wavelengths from the $U-$band to the infrared with
{\it Spitzer}, which generally locate the Lyman-break through the
use of deep $U-$band data, or the Balmer and 4000\AA\ breaks 
through infrared+optical
filters.  However, it has never been shown that a complete
sample of massive galaxies can be selected through these methods and
it remains possible, or even likely, that many massive galaxies are 
missed by not
having a colour or magnitude which fits the criteria being selected
for (see \S 2.2).  For example, ultra-dusty galaxies would possibly have
spectral energy distributions that would not be included in our selection.

As discussed in \S 2.2, the methods for galaxy selection
that we use in this paper for our primary target selection include
the BzK method (Daddi et al. 2004), the IERO method (Yan et al. 2004), the
DRG method (Papovich et al. 2006), and the BX/BM method
(Reddy et al. 2008).   In this section we give a description of the 
relationship between these different methods for determining the population of
massive galaxies at redshifts $z > 2$.  This has been done previously for other
populations at high redshift by Reddy et al. (2005) and Grazian et al. (2007).

A graphical summary of the distribution in redshift for the colour
selection methods for our massive
galaxies is shown in Figure~8, and the distribution of stellar masses 
in Figure~9 using the initial selection described in \S 2.2.   As can
be seen in these figures, there is a slight, but insignificant, difference 
in the redshifts, and stellar mass selection for these massive
galaxies.

We find that the BzK-selected massive galaxies tend to lie towards the
lower range of redshifts, with an average in 
GOODS-N of $\langle z \rangle$ = 2.12$\pm$0.28 and in the GOODS-S of 
$\langle z \rangle$ = 2.17$\pm$0.33.  On the other hand the
IEROs have a higher average redshift range, with 
$\langle z \rangle$ = 2.24$\pm$0.28 in the North and $\langle z \rangle$ = 
2.29$\pm$0.34 in the south.  The DRGs tend to be 
selected with even higher redshifts than either the
BzKs and IEROs, with an average value of $\langle z \rangle = 2.32\pm0.29$ in 
GOODS-S. However, we find that all three methods find galaxies of similar mass,
with the average stellar mass for each type 
$\langle M_{*} \rangle \sim 2 \times 10^{11}$
\solm, and all methods give a similar relatively large range in redshifts.

Furthermore, as can be seen by the different symbols   
in Figure~9, there are many massive systems which are selected by
more than one method.  In fact, we find that nearly all selection
methods overlap with another for some galaxies.   Only a small
fraction of our systems are selected by just one method, with the
IERO selection being the most likely method for finding unique 
galaxies samples.

\begin{figure*}
 \vbox to 110mm{
\includegraphics[angle=0, width=180mm]{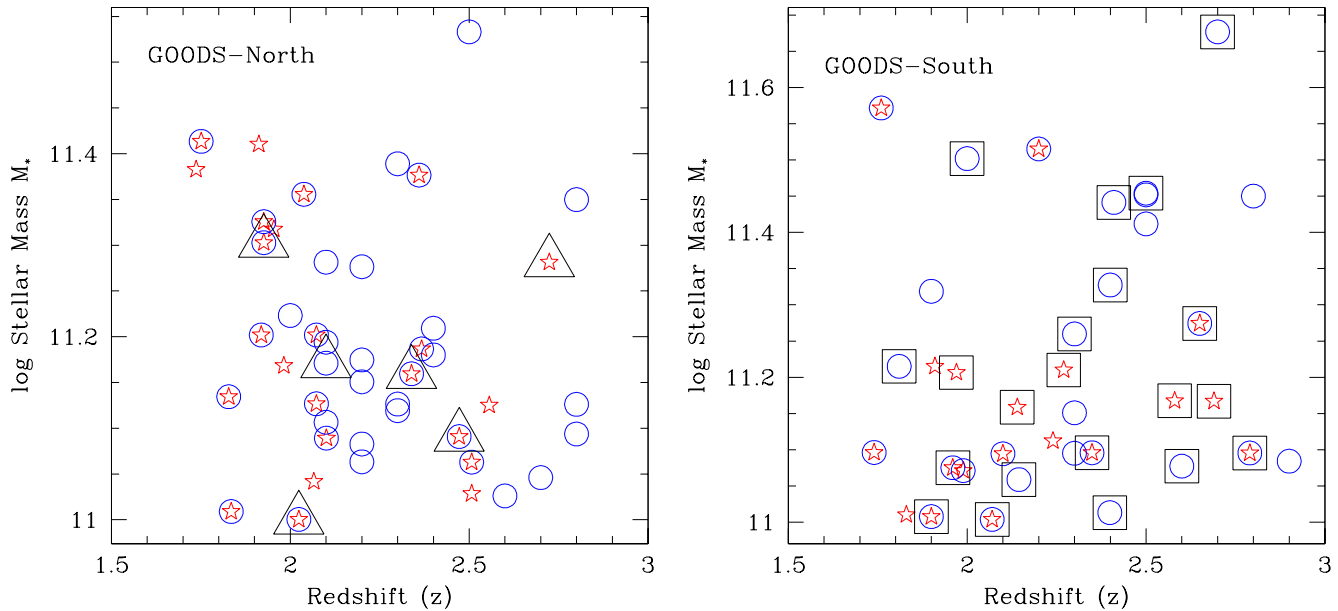}
 \caption{The distribution of our initial
sample of BzKs, IEROs, DRGs, and BM/BX objects from which our
selection of NICMOS fields was made.  The left panel
shows the distribution within GOODS-N, while the
right panel shows this within GOODS-S.  The 
symbols are: stars are BzKs; open circles are IEROs; 
boxes are DRGs which are only found in GOODS-S; and 
triangles are BM/BX objects, which are only located within 
GOODS-N.}
} \label{sample-figure}
\end{figure*}

The breakdown of our selections is such that over all galaxies in
GOODS-N, 24 objects, or $53\pm11$ percent of the systems are selected by the
BzK method.  The corresponding number is 18 systems, or 51$\pm$12 percent
of those in the GOODS-S field.  The IERO selection is the most efficient
for identifying our massive galaxy sample.  In the GOODS-North we find
that 37 galaxies or 82$\pm0.14$ percent are located as  IEROs, while in 
GOODS-S, with 27 systems the fraction is 66$\pm0.15$ percent.  
In the GOODS-S where
we are able to use the DRG method, we find that a total of 21 galaxies
in our massive galaxy sample are selected, or 77$\pm0.14$ percent.  Finally,
we note that only a small fraction of our sample of massive galaxies in
the GOODS-North
at $1.7 < z < 2.9$ are detected through the BM/BX selection method (Reddy
et al. 2008) (Figure~9).
  
There are however a few biases which can produce some of these results.
The first is that the BzK method, as described in \S 2.2, is limited to
$K_{\rm Vega} = 22.0$ and $20.5$ in the GOODS South and North, respectively.
The method will not find bluer galaxies near $z \sim 3$, which will drop out
of the sample at around $z \sim 2.5$. However according to Daddi et al. 
(2004) (eq. 6) we will be largely complete in mass of M$_{*} > 10^{11}$ \solm
at $z < 2.5$ -- through most of our redshift range.  Furthermore, the IERO
selection is potentially the most successful in this experiment due to
the depth of the IRAC imaging compared to the K-band data used in the
DRG and BzK methods. Furthermore, the blunt colour selection for the IEROs
will find more objects, although it remains possible that there are more
false positives with this method of colour selection.
We however conclude that no one single method for locating
distant massive galaxies can be used to find complete samples and that
either a combination of different methods, or a photometric redshift
selected sample, is essential.  

\subsection{Stellar Mass Distribution}

In this section we investigate the stellar masses which we compute
based on our $H_{160}-$band detections matched with our ACS imaging.  These
stellar masses are computed as described in \S 3.3.  This allows us
to examine both the distribution of the stellar masses which we
measure for the GNS sample, as well as test the differing methods outlined
in \S 2.2 for selecting high-redshift galaxies.  These stellar masses
will be the focus of a detailed analysis in Mortlock et al. (2010).

We present in Figure~7 the stellar mass distribution of our
sample out to $z \sim 5$. We also show on this figure the stellar
mass of a maximally old stellar population which would still be
detected at each redshift. Note that there is a slight gap near
redshift $z \sim 1.8$ which is likely partially the result of
photometric redshift systematic errors.   This gap is however also
quite small roughly $\delta z = 0.05$ in size, and there is no
dependence of stellar mass or color in the galaxies that are within this
gap.

We use these stellar masses to determine the completeness and ability
of colour selection methods to find the highest mass galaxies at 
$z > 2$, and to measure their masses accurately.  As we are using
a heterogeneous selection for our high-mass sample, it is important
to carry out this comparison to determine how and whether a stellar
mass selected sample would be similar, and if not, then in what
way it is different.

We find in our new photo$-z$/mass $H_{160}$-based catalogue, independent of our 
original colour selection catalogue (\S2.2), that between $1.7 < z < 2.9$ 
there are 75 massive
galaxies within our criteria of $M_{*} > 10^{11}$ \solm, using our
new $BVizH$ stellar masses and photometric redshifts.  Nearly
all of these galaxies are selected by the colour methods for finding high
redshift massive galaxies (\S 2.2), thereby showing that we have a nearly
complete sample of massive galaxies at high redshift, sans systems
that are extremely dusty that would not be measured with our
photometric redshifts accurately. The reason
there is a slight difference from our 80 original galaxies is that
the computation of photometric redshifts and stellar masses from
our $H_{160}-$band selected catalog are slightly different from those which
were used to construct the original catalogue.  Overall, if we consider
a slightly wider stellar mass and redshift range, we are able
to recover all but ten of the systems which were originally suggested
by our initial colour selected analysis (\S 2.2) to be within our stellar mass 
and redshift range of interest.  Although we find that some galaxies are not
selected by our methods, these are likely to be systems which
just missed our initial criteria, based on our strict
stellar mass and photometric redshift cut.

Overall, we find that the average stellar mass difference between our 
originally estimated masses,
from the colour selected samples, and our our new calculations, is 0.03 dex, 
with a larger scatter of
0.39 dex.  This is slightly larger than our 0.2 dex random error 
measurement, and this is the result of differing redshifts between
the two samples.  This
difference is furthermore reduced to 0.29 dex in scatter when
we examine galaxies for which the two redshift estimates are within $\delta z = 0.1$ 
of their pre- and post-redshift measures.

In other words, when we examine only those galaxies for
which both redshift estimates are near to each other, we find a much
smaller different between the measured stellar masses.  The differences
in the stellar masses can be explained therefore by the
fact that the redshifts are different between the two samples.  We furthermore find 
that the photometric redshift measures for our massive galaxy sample 
from our original catalog compared with the new measures from the 
$H_{160}-$band selected sample are $\delta z/(1+z) = 0.05$, similar to the 
quality of our overall photo-z quality when compared to the  measured 
spectroscopic redshifts.

\subsection{Properties of Massive Galaxies at $z > 2$}

\subsubsection{Previous Investigations}

One of the major focuses of the GNS is to examine the
properties of massive galaxies at $z > 2$.  In the
past this type of analysis
has generally been performed at lower redshifts, at $z < 2$, where
properties of the massive galaxy population are now well
described (e.g., Conselice et al. 2007; Trujillo et al. 2007;
Bundy et al. 2008; Foucaud et al. 2010).  Examining the galaxy 
population at higher redshifts is more challenging due to the fact
that spectroscopic redshifts are difficult to obtain for
a sizeable fraction of galaxies.  However, some early attempts have
been performed which suggest that significant information is
obtainable through deep spectroscopy of distant massive galaxies
(e.g., Kriek et al. 2008).

Despite the lack of spectroscopic redshifts for our sample, we can still make
progress using photometric redshifts and stellar mass measures,
which have already been used in many papers for understanding the
evolution of the massive galaxy population at higher redshifts.  
While we will not providing a detailed analysis of the massive
galaxy population at $z > 2$ within this paper, we give some
basic features, as well as provide information that will
be used in other
papers that will follow this one in terms of the analysis
of these distant galaxies.

We have previously published an analysis of the
size evolution of massive galaxies using this same
data (e.g., Buitrago et al. 2008),
finding that the sizes of our massive galaxy sample are much
smaller by a factor of two to five compared with similar stellar mass
selected galaxies in today's Universe.  Tables~5 and 6 list the
morphological properties of our original 80 galaxy sample.  This include
the GALFIT values of the Sersic index ($n$) and the 
effective radii ($R_{\rm e}$), as well as other shape measures such
as the position angle (P.A.), axis ratio (b/a), effective surface
brightness, as well as the photometric and spectroscopic redshifts for
these systems.

Using this data set combined with previous work at lower
redshifts (e.g., Trujillo et al. 2007; Cassata et al. 2010) 
that there is a gradual
increase in the sizes of galaxies when viewed at lower redshifts.
Understanding how these galaxies become larger at lower redshifts 
is one of the 
primary focuses of studies of massive galaxies, and a solution to
this problem remains outstanding.

\vspace{1cm}
\setcounter{table}{2}
\begin{table*}
\begin{minipage}{160mm}

{\tiny \caption{Basic information and photometry  for our initial colour selected Sample of galaxies with $M_{*} > 10^{11}$ \solm within the GOODS-North Field}
 \label{tab1}
 \begin{tabular}{@{}cccccccccccr}
  \hline
\hline
ID & Type & RA (J2000) & Dec (J2000) & z$_{\rm phot}$ & M$_{*}$ ($\times 10^{11} M_{\odot}$) & $B_{\rm 450}$ & $V_{\rm 606}$ & $i_{775}$ & $z_{850}$ & $H_{160}$ \\
\hline
43 & 4   & 189.1255493  &  62.115509 &  2.20 &  1.21 &  27.05$\pm$0.32 &  26.82$\pm$ 0.20 &  26.04$\pm$0.16  & 25.44$\pm$0.10 & 22.60$\pm$0.10 \\
77 & 1 &  189.132522 &  62.112205 &  1.91 &  2.57 &  27.82$\pm$0.86 &  26.89$\pm$0.27 &  25.76$\pm$0.14 &  25.01$\pm$0.08 &  22.07$\pm$0.06 \\  
21 &  4 &  189.135406 &  62.117168 &  2.70 &  1.11 &  26.16$\pm$0.07 &  25.60$\pm$0.03 &  25.35$\pm$0.03 &  25.34$\pm$0.04 &  23.96$\pm$0.16 \\  
227 &  5 &  189.119278 &  62.135971 &  2.07 &  1.59 &  26.60$\pm$0.13 & 26.48$\pm$0.10 &  26.58$\pm$0.17 &  26.32$\pm$0.15 &  25.23$\pm$0.26 \\  
373 &  5 &  189.058517 &  62.163517 &  2.50 &  1.16 &  ...            &  ...           &  ...            & ...             &  22.80$\pm$0.07 \\  
552 &  5 &  189.077056 &  62.151042 &  1.92 &  2.12 &  24.92$\pm$0.09 & 24.44$\pm$0.06 &  23.61$\pm$0.04 &  23.10$\pm$0.03   &  21.08$\pm$0.03 \\  
730 &  5 &  189.251312 &  62.152904 &  2.47 &  1.23 &  ...              &  ...           &  ...            & ...               &  24.35$\pm$0.15 \\  
840 &  5 &  189.173446 &  62.167392 &  1.92 &  2.01 &  27.30$\pm$0.33 & 26.30$\pm$0.11 &  25.65$\pm$0.0908 &  25.02$\pm$0.06 &  22.15$\pm$0.06 \\  
856 &  1 &  189.178649 &  62.166355 &  1.74 &  2.41 &  28.24$\pm$0.96 & 26.87$\pm$0.23 &  25.66$\pm$0.1146 &  24.83$\pm$0.06 &  21.77$\pm$0.05 \\  
999 &  1 &  189.142868 &  62.233570 &  1.98 &  1.47 &  26.23$\pm$0.17 & 25.17$\pm$0.07 &  24.27$\pm$0.0463 &  23.38$\pm$0.02 &  21.56$\pm$0.06 \\  
1144 &  5 &  189.503555 &  62.270061 &  2.07 &  1.34 &  28.80$\pm$0.65 & 28.69$\pm$0.47 &  ...             &  26.99$\pm$0.18 &  22.29$\pm$0.08 \\  
1129 &  5 &  189.507492 &  62.271797 &  2.37 &  1.54 &  29.34$\pm$0.77 & 28.28$\pm$0.23  &  27.60$\pm$0.20 &  27.65$\pm$0.23 &  23.21$\pm$0.13 \\  
1257 &  5 &  189.226257 &  62.292339 &  2.02 &  1.00 &  29.64$\pm$2.00 &  29.03$\pm$1.03 &  26.57$\pm$0.17 &  26.27$\pm$0.14 &  23.13$\pm$0.09 \\  
1394 &  5 &  189.399597 &  62.345371 &  2.04 &  2.27 &  27.56$\pm$0.43 &  27.57$\pm$0.37 &  26.90$\pm$0.33 &  25.71$\pm$0.12 &  22.26$\pm$0.06 \\  
1533 &  1 &  189.305038 &  62.179489 &  2.56 &  1.33 &  25.86$\pm$0.12 &  25.57$\pm$0.08 &  25.13$\pm$0.08 &  25.10$\pm$0.09 &  24.38$\pm$0.19 \\  
1666 &  5 &  189.256576 &  62.196266 &  2.36 &  2.38 &  27.61$\pm$0.58 &  26.90$\pm$0.27 &  25.77$\pm$0.14 &  25.30$\pm$0.10 &  21.50$\pm$0.05 \\  
1768 &  1 &  189.273559 &  62.187240 &  1.95 &  2.08 &  28.39$\pm$0.95 &  27.31$\pm$0.28 &  26.68$\pm$0.25 &  25.53$\pm$0.09 &  22.20$\pm$0.12 \\  
1826 &  4 &  189.073135 &  62.261402 &  2.20 &  1.89 &  25.34$\pm$0.06 &  25.21$\pm$0.05 &  25.09$\pm$0.07 &  24.64$\pm$0.05 &  22.60$\pm$0.06 \\  
1942 &  1 &  189.277557 &  62.254707 &  2.51 &  1.07 &  25.07$\pm$0.05 &  24.61$\pm$0.03 &  24.35$\pm$0.04 &  24.07$\pm$0.03 &  22.15$\pm$0.05 \\  
2066 &  4 &  189.300201 &  62.203414 &  2.80 &  2.24 &  26.54$\pm$0.19 &  26.83$\pm$0.19 &  25.79$\pm$0.12 &  25.66$\pm$0.12 &  23.21$\pm$0.13 \\  
2083 &  1 &  189.312072 &  62.201652 &  2.72 &  1.91 &  27.77$\pm$0.45 &  27.65$\pm$0.30 &  26.79$\pm$0.22 &  25.99$\pm$0.11 &  22.83$\pm$0.09 \\  
2049 &  4 &  189.312988 &  62.204704 &  2.40 &  1.62 &  ...            &  ...            &    ...          &  ...            &  23.28$\pm$0.22 \\  
2282 &  4 &  189.306976 &  62.262676 &  2.30 &  1.34 &  28.70$\pm$0.71 &  28.12$\pm$0.34 &  27.95$\pm$0.41 &  26.98$\pm$0.18 &  23.57$\pm$0.11 \\
2411 &  4 &  189.047927 &  62.176132 &  2.10 &  1.48 &  27.61$\pm$0.44 &  26.44$\pm$0.13 &  26.05$\pm$0.15 &  25.54$\pm$0.10 &  22.68$\pm$0.09 \\  
2564 &  5 &  189.210907 &  62.248912 &  1.83 &  1.36 &  ...            &  27.64$\pm$0.40 &  26.46$\pm$0.22 &  25.13$\pm$0.07 &  22.26$\pm$0.07 \\  
2734 &  4 &  189.042160 &  62.146274 &  2.60 &  1.06 &  30.15$\pm$1.85 &  28.02$\pm$0.22 &  27.76$\pm$0.31 &  27.25$\pm$0.21 &  23.83$\pm$0.13 \\  
2678 &  4 &  189.047424 &  62.148479 &  2.50 &  3.41 &  ...              &  27.23$\pm$0.29 &  26.36$\pm$0.23 &  25.50$\pm$0.11 &  22.18$\pm$0.05 \\  
2764 &  4 &  189.052475 &  62.143322 &  2.20 &  1.42 &  26.70$\pm$0.17 &  26.21$\pm$0.09 &  25.60$\pm$0.09 &  25.20$\pm$0.07 &  22.97$\pm$0.12 \\  
2902 &  4 &  189.091293 &  62.267700 &  2.00 &  1.67 &  28.33$\pm$0.77 &  27.16$\pm$0.22 &  26.53$\pm$0.20 &  25.95$\pm$0.13 &  23.17$\pm$0.14 \\  
2837 &  4 &  189.094375 &  62.275016 &  2.30 &  2.45 &  25.16$\pm$0.08 &  25.02$\pm$0.06 &  24.62$\pm$0.07 &  24.26$\pm$0.05 &  22.68$\pm$0.07 \\  
2965 &  4 &  189.079818 &  62.244968 &  2.80 &  1.34 &  29.24$\pm$2.75 &  26.26$\pm$0.13 &  25.11$\pm$0.07 &  25.14$\pm$0.08 &  23.06$\pm$0.11 \\  
3036 &  4 &  189.087020 &  62.237724 &  2.10 &  1.28 &  25.37$\pm$0.10 &  24.93$\pm$0.06 &  24.63$\pm$0.06 &  24.01$\pm$0.04 &  22.34$\pm$0.07 \\  
3126 &  5 &  189.130340 &  62.166198 &  2.10 &  1.23 &  27.50$\pm$0.44 &  26.44$\pm$0.13 &  26.00$\pm$0.15 &  25.29$\pm$0.09 &  23.18$\pm$0.10 \\  
3250 &  4 &  189.229095 &  62.138568 &  2.30 &  1.32 &  26.34$\pm$0.12 &  26.44$\pm$0.09 &  26.01$\pm$0.10 &  26.04$\pm$0.11 &  23.11$\pm$0.11 \\  
3422 &  4 &  189.280883 &  62.344234 &  2.80 &  1.24 &  28.26$\pm$0.51 &  26.82$\pm$0.14 &  26.69$\pm$0.17 &  26.47$\pm$0.15 &  23.79$\pm$0.19 \\  
3387 &  5 &  189.294021 &  62.347286 &  1.84 &  1.02 &  28.65$\pm$1.15 &  27.52$\pm$0.29 &  26.53$\pm$0.16 &  25.75$\pm$0.09 &  22.73$\pm$0.08 \\  
3582 &  4 &  189.098754 &  62.169300 &  2.40 &  1.51 &  ...            &  ...            & ...             & ...             &  23.68$\pm$0.20 \\
3629 &  4 &  189.182952 &  62.272567 &  2.10 &  1.91 &  26.56$\pm$0.19 &  25.44$\pm$0.05 &  24.88$\pm$0.05 &  24.40$\pm$0.04 &  21.82$\pm$0.06 \\  
3818 &  5 &  189.202041 &  62.317256 &  1.75 &  2.59 &  28.26$\pm$1.36 &  26.47$\pm$0.25 &  25.28$\pm$0.13 &  24.15$\pm$0.05 &  21.52$\pm$0.05 \\  
3766 &  4 &  189.205612 &  62.322628 &  2.10 &  1.56 &  26.10$\pm$2.00 &  25.44$\pm$0.08 &  24.86$\pm$0.08 &  24.51$\pm$0.06 &  22.14$\pm$0.06 \\  
3822 &  4 &  189.219863 &  62.316909 &  2.20 &  1.16 &  ...            &  29.61$\pm$1.19 &  28.48$\pm$0.62 &  27.36$\pm$0.24 &  23.46$\pm$0.09 \\  
3970 &  5 &  189.331710 &  62.205925 &  2.34 &  1.44 &  27.66$\pm$0.43 &  27.02$\pm$0.18 &  26.15$\pm$0.13 &  25.87$\pm$0.11 &  23.00$\pm$0.10 \\  
4121 &  5 &  189.456344 &  62.233276 &  1.92 &  1.59 &  ...            &  26.29$\pm$0.08 &  25.11$\pm$0.04 &  24.70$\pm$0.03 &  21.92$\pm$0.08 \\  
4033 &  1 &  189.464111 &  62.244133 &  2.07 &  1.10 &  24.82$\pm$0.07 &  24.54$\pm$0.06 &  23.86$\pm$0.05 &  23.68$\pm$0.05 &  21.61$\pm$0.04 \\  
4239 &  4 &  188.981262 &  62.173790 &  2.20 &  1.50 &  26.47$\pm$0.22 &  25.97$\pm$0.11 &  25.70$\pm$0.14 &  25.05$\pm$0.09 &  22.99$\pm$0.13 \\   
\hline
\end{tabular} \\
}
{\small Notes. The ID for each object is from our final NIC3 catalog of objects, the column 'Type' refers to the colour selection method in which 
these galaxies were found, with 1=BzK, 
4=IERO, 5=BzK+IERO; $z_{\rm phot}$ is the initial measured photometric redshift; Mass is the stellar mass in units of $10^{11}$ \solm.  The
remaining panels give our $BVizH$ photometry for these systems.}
\end{minipage}
\end{table*}

\begin{figure}
 \vbox to 120mm{
\includegraphics[angle=0, width=90mm]{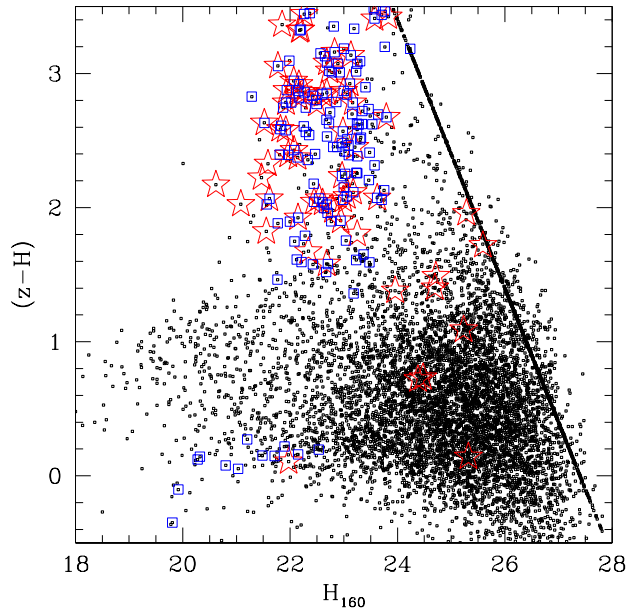}
 \caption{The $H_{160}-$band vs. $(z-H)$ colour diagram for our $H_{160}-$band selected
sample.  Shown here, as red open stars, are the locations of our initial stellar mass
selected sample.  As can be seen, many of the
massive galaxies in our sample are quite red.  The apparent line on the
right of the diagram is the result of galaxies which are undetected in the
z-band.}
} \label{sample-figure}
\end{figure}

\begin{figure}
 \vbox to 140mm{
\includegraphics[angle=0, width=90mm]{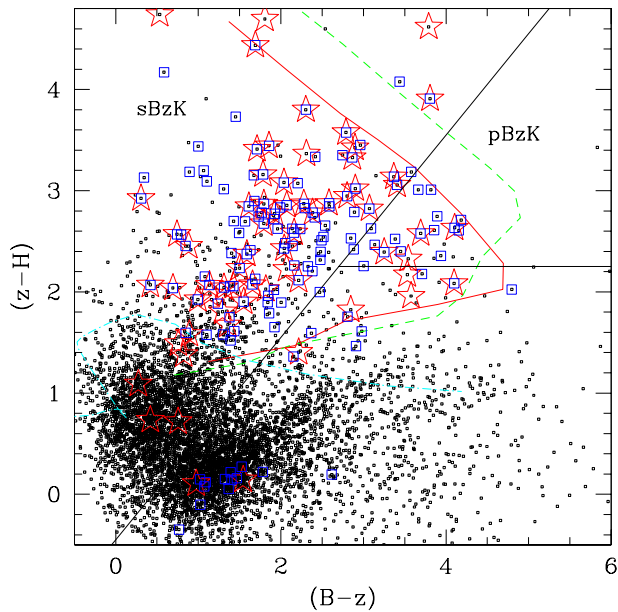}
 \caption{The $(z-H)$ vs. $(B-z)$ diagram for the $H_{160}-$band selected
galaxies in our sample. The red stars show the location of our initial
massive galaxy sample based on colour selections (\S 2.2), while the blue
squares show the location of massive systems with $M_{*} > 10^{11}$ \solm
at $1.8 < z < 3.0$ identified through photometric redshifts.  The various 
lines on this figure show the change in colour for
simple single stellar populations of a given age as observed at various redshifts.
Changes in colour seen for these models are purely due to redshift effects.
The cyan dot-dashed line shows the evolution of a 100 Myr  single stellar
population from redshifts of $z = 0$ to $z = 3.8$.
The green dashed line shows a 2.5 Gyr single stellar population as observed at redshifts
from $z = 0$ to $z = 2.6$, while the red solid line shows a 11 Gyr old stellar population
as viewed from $z = 0$ to $z = 2.2$.}
\vspace{8cm}
} \label{sample-figure}
\end{figure}

\vspace{1cm}
\setcounter{table}{3}
\begin{table*}
\begin{minipage}{160mm}
{\tiny \caption{Basic information and photometry  for our initial colour selected sample of galaxies with $M_{*} > 10^{11}$ \solm within the GOODS-South Field}
 \label{tab1}
 \begin{tabular}{@{}cccccccccccc}
  \hline
\hline
ID & Type & RA (J2000) & Dec (J2000) & z$_{\rm phot}$ & Mass ($\times 10^{11} M_{\odot}$) & $B_{\rm 450}$ & $V_{\rm 606}$ & $i_{775}$ & $z_{850}$ & $H_{160}$ \\
\hline
4299 &  6 &  53.0918007 &  $-$27.8028107 &  2.40 &  2.13 &  27.01$\pm$0.19 &  25.78$\pm$0.06 &  25.65$\pm$0.08 &  25.47$\pm$0.08 &  25.32$\pm$0.22 \\  
4281 &  6 &  53.0938988 &  $-$27.8011951 &  2.60 &  1.20 &  25.63$\pm$0.05 &  25.49$\pm$0.04 &  25.40$\pm$0.07 &  25.21$\pm$0.06 &  24.48$\pm$0.14 \\  
4348 &  3 &  53.1008377 &  $-$27.8082333 &  1.97 &  1.61 &  ...            &  29.03$\pm$0.62 &  27.20$\pm$0.20 &  27.24$\pm$0.23 &  25.28$\pm$0.28 \\  
4434 &  6 &  53.0976601 &  $-$27.7153015 &  2.14 &  1.15 &  25.80$\pm$0.07 &  25.18$\pm$0.04 &  24.80$\pm$0.04 &  24.49$\pm$0.03 &  22.44$\pm$0.05 \\  
4399 &  4 &  53.1008263 &  $-$27.7117653 &  2.30 &  1.25 &  26.96$\pm$0.18 &  26.47$\pm$0.10 &  26.24$\pm$0.14 &  26.20$\pm$0.15 &  24.71$\pm$0.23 \\  
4557 &  3 &  53.0891876 &  $-$27.7600765 &  2.27 &  1.62 &  27.40$\pm$0.51 &  26.88$\pm$0.29 &  25.93$\pm$0.20 &  24.82$\pm$0.08 &  21.94$\pm$0.08 \\  
4754 &  6 &  53.1201096 &  $-$27.8082657 &  2.00 &  3.18 &  29.09$\pm$1.06 &  28.38$\pm$0.51 &  27.14$\pm$0.28 &  26.30$\pm$0.14 &  22.73$\pm$0.08 \\  
4706 &  7 &  53.1231232 &  $-$27.8033943 &  2.34 &  1.25 &  28.06$\pm$0.54 &  26.44$\pm$0.09 &  25.63$\pm$0.07 &  24.99$\pm$0.04 &  22.16$\pm$0.08 \\  
4882 &  5 &  53.1717033 &  $-$27.8256683 &  1.74 &  1.25 &  27.61$\pm$0.40 &  26.34$\pm$0.11 &  25.14$\pm$0.06 &  24.36$\pm$0.04 &  21.97$\pm$0.05 \\  
4941 &  1 &  53.2300110 &  $-$27.8507748 &  1.83 &  1.02 &  26.87$\pm$0.17 &  26.38$\pm$0.10 &  26.14$\pm$0.13 &  26.34$\pm$0.17 &  21.60$\pm$0.05 \\  
5171 &  6 &  53.0632668 &  $-$27.6996498 &  2.39 &  1.03 &  26.39$\pm$0.12 &  25.56$\pm$0.05 &  25.04$\pm$0.05 &  24.84$\pm$0.05 &  22.94$\pm$0.08 \\  
5281 &  5 &  53.0859909 &  $-$27.7091026 &  2.10 &  1.24 &  ...            &  28.37$\pm$0.35 &  27.33$\pm$0.22 &  27.31$\pm$0.25 &  25.59$\pm$0.26 \\ 
5445 &  6 &  53.1245880 &  $-$27.8932495 &  2.50 &  2.85 &  ...            &  ...            & ...             &  ....           & 23.54$\pm$0.12 \\  
5372 &  4 &  53.1255684 &  $-$27.8864536 &  2.90 &  1.21 &  27.35$\pm$0.24 &  26.14$\pm$0.08 &  25.77$\pm$0.10 &  25.69$\pm$0.10 &  23.62$\pm$0.15 \\  
5533 &  7 &  53.1289139 &  $-$27.9036846 &  2.79 &  1.25 &  ...            &  ...            &  ...            &  ...            &  23.60$\pm$0.11 \\  
5524 &  3 &  53.1332588 &  $-$27.9029388 &  2.58 &  1.47 &  ...            &  26.22$\pm$0.06 &  25.20$\pm$0.04 &  25.09$\pm$0.04 &  22.17$\pm$0.07 \\  
5764 &  7 &  53.2252083 &  $-$27.8738060 &  2.65 &  1.88 &  ...            &  28.23$\pm$0.46 &  27.21$\pm$0.29 &  26.50$\pm$0.17 &  22.66$\pm$0.07 \\  
5853 &  6 &  53.0508537 &  $-$27.7137222 &  2.41 &  2.76 &  26.35$\pm$0.20 &  25.70$\pm$0.09 &  25.61$\pm$0.15 &  25.05$\pm$0.11 &  23.25$\pm$0.15 \\  
5933 &  6 &  53.0542488 &  $-$27.7216587 &  2.30 &  1.82 &  ...            &  28.94$\pm$0.47 &  28.00$\pm$0.33 &  27.54$\pm$0.26 &  23.32$\pm$0.11 \\  
6035 &  4 &  53.0555954 &  $-$27.8740025 &  1.90 &  2.08 &  27.52$\pm$0.00 &  26.98$\pm$0.23 &  25.73$\pm$0.13 &  25.21$\pm$0.10 &  21.85$\pm$0.06 \\ 
6114 &  1 &  53.0656776 &  $-$27.8788643 &  2.24 &  1.30 &  28.19$\pm$0.66 &  26.20$\pm$0.10 &  25.57$\pm$0.09 &  24.50$\pm$0.04 &  21.92$\pm$0.05 \\  
6220 &  7 &  53.0717087 &  $-$27.8436356 &  1.90 &  1.02 &  27.19$\pm$0.37 &  26.79$\pm$0.19 &  25.91$\pm$0.14 &  25.27$\pm$0.09 &  22.50$\pm$0.07 \\  
6352 &  7 &  53.0773201 &  $-$27.8595829 &  1.96 &  1.19 &  ...            &  28.90$\pm$0.62 &  28.65$\pm$0.83 &  27.00$\pm$0.21 &  22.37$\pm$0.08 \\  
6468 &  4 &  53.1385193 &  $-$27.6717854 &  2.80 &  2.82 &  25.92$\pm$0.18 &  26.12$\pm$0.16 &  25.25$\pm$0.12 &  24.78$\pm$0.08 &  22.80$\pm$0.10 \\  
6584 &  5 &  53.0260849 &  $-$27.6909122 &  1.99 &  1.18 &  26.54$\pm$0.0 &  27.42$\pm$0.56 &  25.95$\pm$0.25 &  24.50$\pm$0.09 &  22.07$\pm$0.09 \\  
6575 &  4 &  53.0354462 &  $-$27.6900806 &  2.50 &  2.58 &  25.03$\pm$0.07 &  26.12$\pm$0.16 &  25.53$\pm$0.16 &  24.61$\pm$0.08 &  22.54$\pm$0.12 \\  
6876 &  4 &  53.0400429 &  $-$27.6852055 &  2.50 &  2.83 &  26.30$\pm$0.00 &  27.08$\pm$0.27 &  25.98$\pm$0.17 &  25.55$\pm$0.15 &  22.98$\pm$0.16 \\  
7090 &  6 &  53.0578766 &  $-$27.8335018 &  2.70 &  4.75 &  ...            &  ...            &  ...            & ...             &  22.22$\pm$0.06 \\ 
7155 &  3 &  53.1175194 &  $-$27.9107571 &  2.69 &  1.47 &  28.32$\pm$0.73 &  26.84$\pm$0.17 &  26.09$\pm$0.14 &  26.10$\pm$0.16 &  24.70$\pm$0.39 \\  
7320 &  7 &  53.1156578 &  $-$27.8717003 &  2.07 &  1.01 &  ...            &  ...            &     ...         & ...             &  26.36$\pm$0.53 \\  
7425 &  6 &  53.1271477 &  $-$27.8345642 &  1.81 &  1.64 &  27.46$\pm$0.36 &  26.21$\pm$0.15 &  24.91$\pm$0.08 &  23.91$\pm$0.03 &  21.59$\pm$0.06 \\  
7677 &  5 &  53.1830482 &  $-$27.7089996 &  1.76 &  3.73 &  26.32$\pm$0.26 &  25.55$\pm$0.11 &  23.80$\pm$0.04 &  22.79$\pm$0.02 &  20.62$\pm$0.04 \\  
7970 &  4 &  53.0282135 &  $-$27.7788277 &  2.30 &  1.41 &  27.80$\pm$0.52 &  27.56$\pm$0.38 &  26.22$\pm$0.18 &  25.77$\pm$0.14 &  22.69$\pm$0.06 \\  
8140 &  1 &  53.1410255 &  $-$27.7667332 &  1.91 &  1.64 &  25.76$\pm$0.10 &  25.10$\pm$0.06 &  24.45$\pm$0.05 &  23.69$\pm$0.03 &  21.47$\pm$0.05 \\  
8213 &  3 &  53.1628799 &  $-$27.7122879 &  2.14 &  1.44 &  23.05$\pm$0.02 &  22.83$\pm$0.01 &  22.30$\pm$0.01 &  22.07$\pm$0.01 &  21.97$\pm$0.06 \\  
\hline
\end{tabular} \\ 
}
{\small Notes. The ID for each object is from our final NIC3 catalog of objects, the column 
'Type' refers to the colour selection method in which 
these galaxies were found, with 1=BzK, 3=BzK+DRG,
4=IERO, 5=BzK+IERO, 6=DRG+IERO, 7=BzK+DRG+IERO; $z_{\rm phot}$ is the initial measured 
photometric redshift; Mass is the stellar mass in units of $10^{11}$ \solm.  The
remaining panels give our $BVizH$ photometry for these systems.}
\end{minipage}
\end{table*}

Another issue we have used this data set to address is
the merger history of these massive galaxies at $z < 3$ (Bluck
et al. 2009, 2010 in prep).  By investigating pairs of galaxies at these
redshifts, we find that the merger fraction for massive galaxies
contains a steep decline,
which falls as $\sim (1+z)^{3}$, and that overall there are
no more than roughly two or three major mergers occurring for these
massive galaxies at $z < 3$, suggesting that at most the
stellar mass is tripled by such mergers, and that furthermore,
these mergers are probably not directly producing the increase in the
sizes of these massive galaxies (e.g., Buitrago et al. 2008).   
This is further confirmed when examining similar results at lower 
redshifts (Conselice et al. 2009).   We are currently investigating
the merger history of GNS galaxies through the use of the CAS morphological 
parameters using the NICMOS imaging (e.g.,  Conselice
et al. 2008; Bluck et al. 2010).

We are also investigating the star formation history for these
systems by using MIPS photometry and rest-frame UV fluxes (Bauer
et al. 2010; Weinzirl et al. 2010).  Furthermore, we are looking at the 
environmental
properties of our galaxies, and are comparing their environments
to similar mass galaxies at lower redshifts (e.g., Gr\"utzbauch et al.
2010).  We are also investigating the surface brightness
profiles and Kormendy relations and minor merger histories
for these galaxies (Bluck et al. 2010b) and at how, by comparing these 
measurements to simulations of nearby galaxies placed at high redshift, the
evolutionary history of these galaxies can be deciphered (e.g.,
Conselice et al. 2010, in prep.).  Finally, we are investigating the more general
mass-selected population at high redshift through the evolution of
stellar mass functions and colours (e.g., Mortlock et al. 2010).

\subsubsection{Colours}

The colours of our massive galaxies are typically quite red, and most
of them have colours $(z-H) > 1.5$ (Figures~10 and 11) and 
with $H_{160}$ magnitudes $H_{160} \sim 23$ AB.  In Figures~10 and 11 the 
original colour selected
massive sample is shown by the open stars.  The objects within our 
new photometric redshift selection which
are not within the original sample also
fall within this area of red colour space, as denoted by the open 
blue squares (Figure~10 \& 11).
This also shows that these galaxies are largely very faint in
the optical, and that the best way to study them is
in the near-infrared.  However, as can be seen through the blue
squares on Figure~10, some fraction of the systems which are selected
by the photometric redshift technique have bluer colours,  with
flat spectrum colours of $(z-H) \sim 0$.  These galaxies would
not have been selected through our colour techniques due to the
fact that such systems are too blue, and thus are hard to distinguish
from lower redshift galaxies.  However, as we will see, these objects
are not a dominant part of the population, and in fact most of the
global quantities calculated for these massive galaxies at $z > 2$ are
largely the same whichever of these two methods for selection is used.

We can also get some basic idea of the star formation history and stellar 
populations of these massive galaxies by examining their position in 
colour-colour space (Figure~11) and their location within rest-frame
colour-magnitude diagrams (Figure~12).  We only show galaxies
down to $M_{*} = 10^{10}$~\solm on these colour-magnitude diagrams,
where we are complete in our selection of galaxies.    What we find is that our
massive galaxies span a range in ($B-z$) colour, but are
all fairly red in ($z - H$), with values larger than ($z - H$) $= 2$
for most systems.   We find that a large fraction of our massive
galaxy systems are
within the star-forming region of the BzK diagnostic plots (e.g.,
Daddi et al. 2007; Lane et al. 2007), after converting the diagnostics
to a BzH selection using a typical colour of $(H-K) \sim 0.25$ for
a galaxy at $z \sim 2$.  
This implies that within this
selection, there are a significant number of galaxies
with enough observed $B-$band flux to be considered star-forming
systems  (see also Bauer et al. 2010).  The 
lines on Figure~11 furthermore show where various
single stellar populations would lie in this parameter space whose
colours are affect simply by k-corrections. 

We furthermore use the redshifts and magnitudes of our galaxies to
calculate rest-frame $(U-B)$ colours, which are directly compared with
colours from lower redshift galaxies of similar masses (e.g, Conselice
et al. 2007).  The result of this rest-frame $U-B$ vs. $M_{\rm B}$ diagram
is shown in Figure~12, where we have divided up our sample into 
ultra-massive galaxies with $M_{*} >$ \hmass, medium mass galaxies with
\hmass $>$ $M_{*} >$ \mass\, and those systems with more modest,
but still relatively high masses, with \mass $>$ $M_{*} >$ \lmass.  We
also show this evolution in colour-magnitude for our galaxies divided into
different redshift bins.  Note that we are complete for $M_{*} >$ \mass galaxies up to
$z = 3$.  The blue and red lines on Figure~12 show the demarcation between the red
sequence and the blue cloud, and the location of the red sequence as seen at lower 
redshifts (Faber et al. 2007).  These lines are evolved passively from lower
redshift to higher-$z$.

What one can see immediately from Figure~12 is that most of the massive
galaxies, which tend also to be bright, are near or close to the 
red-sequence, while the lower mass galaxies are more often bluer systems,
with in fact, very few of the galaxies with lower masses in our sample 
within the red sequence.  The bi-modality is particularly
present at $2.5 < z < 3$, where nearly all the massive galaxies with
$M_{*} >$ \mass are exclusively on the red sequence, while those less
massive galaxies are found within the blue cloud.  As we go to lower
redshift it appears that some of the massive galaxies are now seen
in the blue cloud.

This apparent evolution from the red sequence to the blue cloud is likely
due to the fact that the number densities of massive galaxies increases
with time, and thus massive galaxies at $1.5 < z < 2.0$ are not
the same type of system found at higher redshifts.  In other words, 
the build-up
of massive galaxies occurs by adding bluer galaxies, that were formerly 
lower mass
earlier, to the high mass bin. Furthermore, as discussed in Bauer et al.
(2010), the fact that these massive galaxies are red does not necessarily
imply that they are `red and dead', but in fact, that they are undergoing
dusty star formation (e.g., Papovich et al. 2006). 
A detailed discussion of this is presented
in Bauer et al. (2010) and Gr\"utzbauch et al. (2010).

\begin{figure*}
 \vbox to 110mm{
\includegraphics[angle=0, width=180mm]{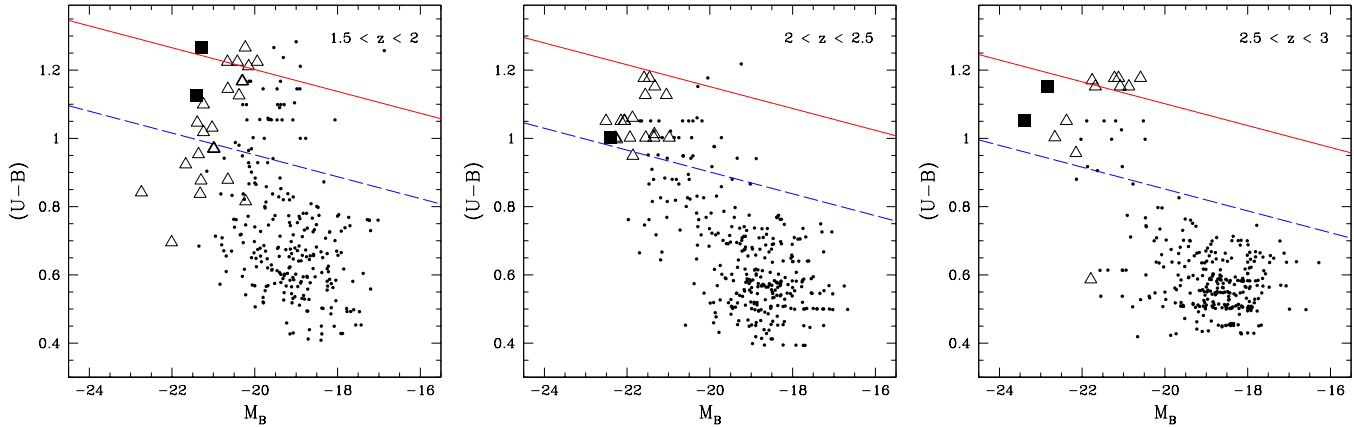}
 \caption{Colour-magnitude diagram in rest-frame units for our
sample of galaxies. Shown are the data plotted in different ways
depending on their stellar mass.  The most massive galaxies
with $M_{*} >$ \hmass are shown as solid squares, those
at \mass $<$ $M_{*} <$ \hmass are shown as open triangles,
and the dots are for those galaxies with stellar masses, 
$M_{*} <$ \mass.  The red solid line is the red sequence as determined
by Faber et al. (2007), while the blue dashed line shows the
separation at which galaxies are considered to be within the blue cloud.
Both of these lines are evolved with redshifts for a passively evolving
stellar population.}
} \label{sample-figure}
\end{figure*}

\subsubsection{The Evolution of Massive Galaxy Number Densities}

\begin{figure}
 \vbox to 120mm{
\includegraphics[angle=0, width=90mm]{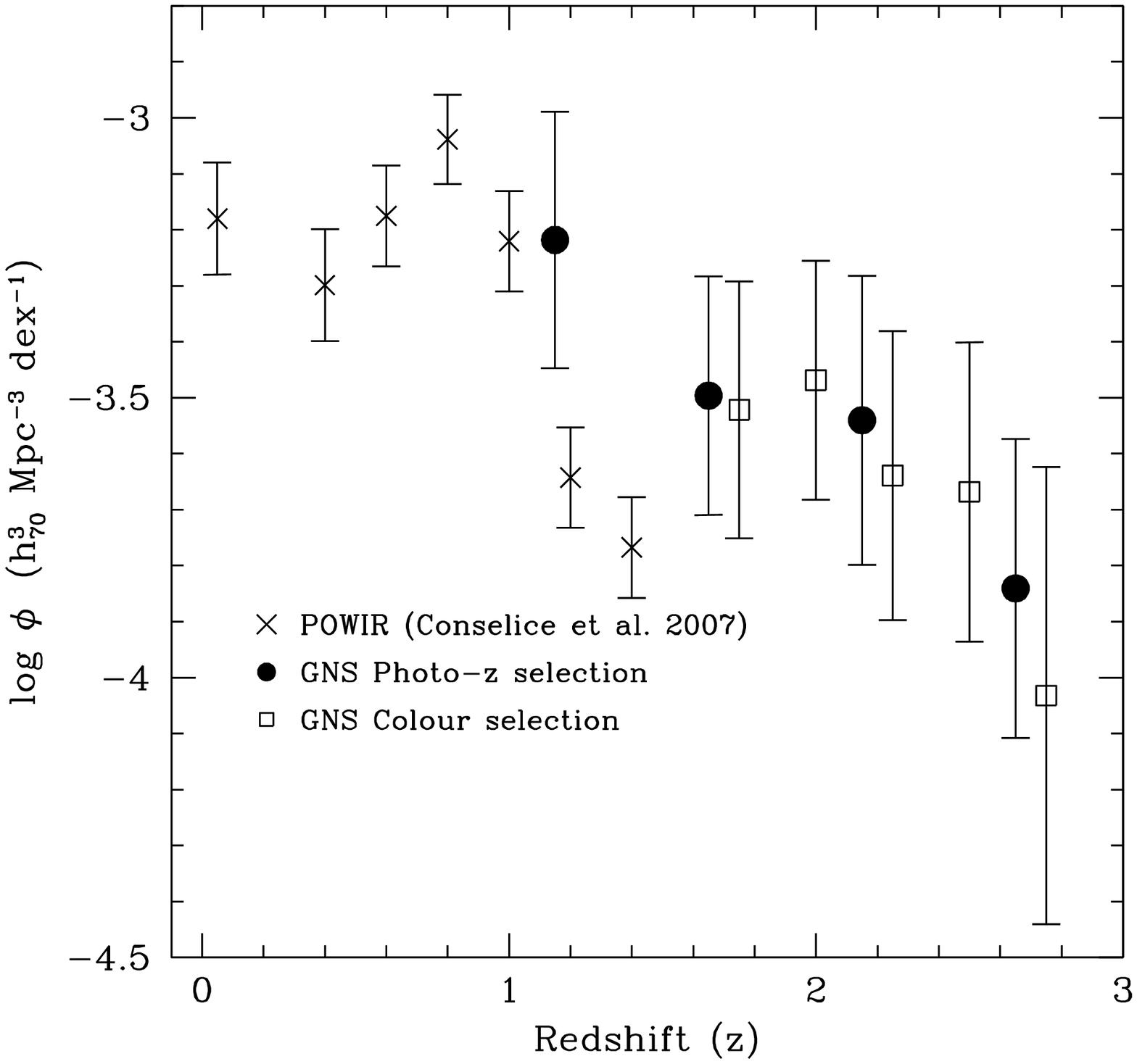}
 \caption{The number density evolution in units of h$_{70}^{3}$ Mpc$^{-3}$ for
galaxies with stellar masses, $M_{*} >$ \mass as seen within the GNS
and within the {\it POWIR} survey (Conselice et al. 2007a).  We show here the
number densities calculated using two different galaxy sample selections,
both a photo$-z$ method and the original colour-selection method, with both
methods showing relatively good agreement.}
} \label{sample-figure}
\end{figure}

One of the major ways to examine the evolution of massive galaxies is to
investigate how their number densities evolve with time
(e.g., Conselice et al. 2007).  Various studies
have previously examined how luminosity and mass-selected samples have evolved
over redshift.  A very popular way to do this is by examining galaxy luminosity
functions, typically in the $B-$band, and to fit the distribution of 
luminosities to a Schechter function, giving a characteristic luminosity
$L^{*}$ as well as a faint-end slope ($\alpha$), and normalization given
by the parameter $\phi$.  Typically one measures evolution in the
galaxy population by fitting these parameters, and then determining how
they have evolved over time.  This is useful for determining how the
global galaxy population changes, or in the case of faint or low
mass galaxies, through examining how the $\alpha$ parameter evolves
with redshift. 

An alternative approach is to examine galaxies based on a certain luminosity
or stellar mass threshold,
such as our selection in this paper of $M_{*} > 10^{11}$ \solm.  There are 
many reasons for
believing that stellar mass is a better indicator for tracing the 
evolution of galaxies than luminosity. One reason is that  
stellar mass measurements in principle do not depend upon the
ongoing star formation rate, whereby for a higher star formation rate 
the galaxy will appear brighter, but whose star formation can be
revealed through its SED, and which can be normalised out.

We have  also found within the GNS, and at $z < 1$ within the {\it POWIR} 
survey
(Conselice et al. 2008), that stellar mass is the most important quantity 
for determining the properties of a galaxy, such as its colour and
star formation rate (e.g., Gr\"utzbauch et al. 2010a,b).    There is a 
strong correlation up to $z \sim 3$, such that we find that galaxies
are redder at high stellar masses than bluer galaxies at lower masses.
This trend stays the same at all redshifts thus far probed, and a similar
trend can be seen when considering the star formation rate and the specific
star formation rate correlations with stellar mass.

Previously,  the number densities of $M_{*} > 10^{11}$~\solm
galaxies at $z < 2$ were investigated using the {\it POWIR/DEEP2} 
data set (Conselice et al. 2007). 
It was found in this previous paper that the number densities of these massive 
galaxies, and more specifically, those of masses \hmass, were largely in place
at $z = 1 - 1.5$, and with a significant number of systems present at
$z = 2$. The fact that these massive galaxies already exist in large
numbers at this redshifts, suggests that we have to go to higher redshifts
to trace the evolution of these systems.  This was one reason for carrying
out the GNS survey, for which we can make an estimate of the number densities
of massive galaxies at even higher redshifts. 

We do this in two ways -- both by using our original colour-selected, and
likely slightly inhomogeneous sample, as well as by the use of our newly
measured photometric redshifts and stellar masses from our $BViz$H
selected sample (Figure~13).  The errors on this plot result from number 
counting statistics only. 

In Figure~13 we
have down-weighted the stellar mass densities at $1.7 < z < 2.9$ to account
for the fact that our fields were selected to contain massive galaxies
with $M_{*} >$ \mass.  The GNS fields were selected based on having at least
one colour selected  massive galaxies at $1.7 < z < 2.9$ 
in the NIC3 field of view within the GOODS South and North.  These fields
were however not over selected for massive galaxies, as the number of 
other objects (drop-outs, sub-mm galaxies, BM/BX galaxies) were used to 
make the final selection, as well as lower mass BzKs, DRGs, and IEROs.

Ultimately the selection used 80 unique $M_{*} > 10^{11}$~\solm galaxies.
We calculate the correction factor for this selection by using the original 
list of massive galaxies with these redshifts and stellar masses from
the colour-selected lists over the entire GOODS fields, and calculate 
how many non-overlapping massive galaxies there are.  We find 92 of these
systems in the North and 83 in the South.  This gives a surface density of 
0.58 arcmin$^{-2}$.  The 80 galaxies within the GNS area provides a surface
density of 1.77 arcmin$^{-2}$ in the NICMOS pointings.    Thus, the ratio of 
these gives the over-density for $M_{*} > 10^{11}$~\solm galaxies at $z > 1.7$ 
and $z < 2.9$ in the NICMOS data, which is a factor of 3.05 which we use
to down-weight the number densities we calculate for  the
$M_{*} > 10^{11}$~\solm galaxies.

Using this, what we find is that the massive galaxy density
is roughly constant and similar to its value at $z \sim 0$ up to $z = 1.5$,
with a decline thereafter.  
There is then a real decline at higher redshifts, such that the number
density for these massive galaxies grows by a factor of eight between
$z = 3$ and $z =1.5$.   We find this is the case for both selection methods
for these massive galaxies.  This demonstrates that this epoch between
$z = 1.5 - 3$ is when a large fraction of massive galaxies become massive,
and thus physical processes are ongoing during this epoch which produces
this increase of almost a factor of ten in the number of massive galaxies
with $M_{*} > 10^{11}$~\solm during this relatively short period of 
$\sim 2$ Gyr.  Future GNS papers will address the physical mechanisms which
are producing this evolution.

\section{Summary}

The GOODS NICMOS Survey (GNS) is a 180 orbit {\it Hubble} Space Telescope 
programme designed to obtain deep NICMOS $H_{160}-$band imaging of over
80 massive $>$ \mass galaxies at $1.7 < z < 2.9$.  The depths reached 
are $H_{160} \sim 26.8$ AB (5 $\sigma$), 
allowing for a range of other science questions to be addressed, 
including examining the lower mass galaxy
population present within the same fields and redshifts.   In this paper we
describe the GNS survey, give information about its field selection,
as well as a description of the types of galaxies we initially select
for field placement, and how this compares to a newer $BVizH$ photo$-z$ 
selection based on the combination of $H_{160}-$band and ACS data.  We 
utilize only these five filters so as to have a high fidelity in 
our photometry quality, depth, and resolution.  Our photo$-z$s are 
in fact very good, with a typical $\delta z/(1+z) \sim 0.1$.  

We also examine in this paper ways to select massive 
galaxies at high redshift $z > 2$ in the absence of a significant
number of spectroscopic redshifts.  A very popular method for
determining a galaxy population at $z > 2$ is through the use of
various colour cuts, e.g., BzK galaxies, DRGs, or Lyman-break
galaxies.  We find overall that no one single colour criterion is
able to account for all massive galaxies, and that there is considerable
overlap between the various methods, many of which find the same galaxies. 
Overall, we find that the BzK, DRG and IERO overlap,
with over 50 percent between any two types. IEROs are the type which most
overlaps the other two, and perhaps provides the most complete
sample of at least the red galaxies selected at higher redshifts.

We show that a photometric redshift selection with a strict narrow
redshift and stellar mass range gives slightly different galaxy populations
than the cumulative colour selection. In particular we find that
there are more blue massive galaxies at $z > 2$ than what is found
through the standard colour cuts. Galaxies absent from each of these
bins are missed as they are just outside the stringent mass and
redshift selection criteria we have imposed. A similar issue 
is present for the colour-selection of high$-z$ galaxies -- there 
is always a population just outside a colour-selection
that has intrinsic properties nearly identical to the galaxies within the
original colour based selection. This 
is especially the case if one considers galaxies selected by stellar mass
and redshift.  We find, however, that this is not a significant limitation,
as we obtain the same number densities and average properties using
a colour selection and a photo$-z$ selection.

We also provide a summary of our understanding of how these massive
galaxies evolve at high redshifts and how they are connected to lower
redshift massive galaxies.  Our conclusions regarding an 
analysis of the massive galaxy
population reveal that major mergers are not adequate for driving
the evolution of massive galaxies at $z < 3$ (Bluck et al. 2009).  
Our conclusions based on an examination of the major merger rate
 at $z\sim 2-3$ (Bluck et al. 2009), and the star formation
 properties (Weinzirl et al. 2010, in prep) of the massive galaxies, 
suggests
 that other mechanisms, such as minor mergers and/or gas accretion
 and subsequent star formation are also needed to produce the increase
 in stellar mass within these galaxies over time.
 Furthermore, the 'disky' nature of a large fraction of the massive
 galaxies at $z\sim 2-3$,  and their star formation properties (Weinzirl
 et al. 2010) raises questions on the importance of major mergers in
 building  such systems up at $z>3$.

 We find that even at $z \sim 2.5$ there is a
broad colour-magnitude bimodality in galaxies, such that the massive systems
are nearly always red and luminous, while lower mass galaxies tend to be much
bluer.  The build up of the red sequence at $z > 2.5$ is further investigated
in Gr\"utzbach et al. (2010).

We are also investigating various 
other properties
of these massive galaxies, including the AGN content and how AGN in massive
galaxies evolves through time (Bluck et al. 2010; Weinzirl et al. 2010), 
including how much energy the AGN inputs 
into the galaxy while it is evolving.  Details of these calculations will be 
presented in
future GNS papers utilising the data from this paper, as well as 
data from other space and ground based telescopes, such as {\it Spitzer}
and {\it Chandra}.

In the future, to make progress with massive galaxy evolution in
the redshift range $z > 2$ will require either a very large number
of reliable photometric redshifts,
that are studied in a statistical sense, or ultimately through 
spectroscopic surveys that will acquire redshifts and other physical
information for distant massive galaxies. This should be possible 
with multi-object
NIR spectroscopy which is now becoming feasible with new instrumentation on
8-10m class telescopes.  Probing massive galaxy formation at higher
redshifts will require even larger telescopes and instruments and 
will likely be a major focus of JWST and the TMT/ELT/GMT era.

The data and catalogs as used in the GNS survey are online at:
{\bf http://www.nottingham.ac.uk/astronomy/gns/}
We thank Will Hartley for providing diagnostic lines of stellar 
populations,  as well as Omar Almaini for discussions on various aspects 
of this analysis.  We thank STFC for support through studentships and 
PDRA funding. 
Support was also provided by NASA/STScI grant HST-GO11082.  S. J. and 
T. W. acknowledges support from the National Aeronautics and Space
Administration (NASA) LTSA grant NAG5-13063, NSF grant AST-0607748,
and $HST$ grant GO-11082 from STScI, which is operated by AURA, Inc.,
for NASA, under NAS5-26555.

\vspace{1cm}
\setcounter{table}{4}
\begin{table*}
\begin{minipage}{160mm}
 \caption{Structural parameters for galaxies with $M_{*} > 10^{11}$ \solm within the GOODS-N field.  }
 \label{tab1}
 \begin{tabular}{@{}cccccccccc}
  \hline
\hline
ID & R${\rm e}$$^{\it a}$ (arcsec) & Sersic $n$$^{\it a}$ & b/a & P.A. (deg) & $\mu_{\rm eff}$ (mag arcsec$^{-2}$) & $z_{\rm spec}$ & sep (arcsec) & $z_{\rm phot}$ \\
\hline
43 &  0.14$\pm$0.00 &  1.67$\pm$0.05 &  0.75 &  50.6 & 17.26 &  ... &  ... &  1.56 \\ 
77 &  0.35$\pm$0.01 &  1.35$\pm$0.03 &  0.75 &  -23.9 &  18.82 &  ... &  ... &  1.71 \\  
21 &  0.14$\pm$0.01 &  1.0$\pm$0.20 &  0.38 &  9.6 &  16.92 &  ... &  ... &  2.47 \\  
227 &  0.25$\pm$0.00 &  2.39$\pm$0.05 &  0.73 &  -41.3 &  18.00 &  ... &  ... &  1.79 \\  
373 &  0.27$\pm$0.01 &  0.61$\pm$0.03 &  0.26 &  -17.3 &  16.63 &  ... &  ... &  2.53 \\  
552 &  0.44$\pm$0.01 &  2.36$\pm$0.05 &  0.82 &  77.2 &  18.57 &  ... &  ... &  1.81 \\  
730 &  0.58$\pm$0.01 &  0.48$\pm$0.02 &  0.47 &  28.2 &  19.03 &  ... &  ... &  2.11 \\  
840 &  0.14$\pm$0.00 &  2.38$\pm$0.09 &  0.49 &  17.4 &  16.63 &  ... &  ... &  2.14 \\  
856 &  0.17$\pm$0.00 &  3.63$\pm$0.10 &  0.71 &  -17.8 &  17.63 &  ... &  ... &  1.82 \\  
999 &  0.20$\pm$0.00 &  1.28$\pm$0.03 &  0.68 &  64.3 &  16.86 &  ... &  ... &  1.47 \\  
1144 &  0.49$\pm$0.01 &  1.35$\pm$0.03 &  0.32 &  -24.0 &  18.58 &  ... &  ... &  2.66 \\  
1129 &  0.32$\pm$0.01 &  0.17$\pm$0.04 &  0.69 &  -8.7 &  18.81 &  ... &  ... &  2.77 \\  
1257 &  0.09$\pm$0.00 &  3.45$\pm$0.20 &  0.73 &  -7.3 &  17.04 &  ... &  ... &  1.66 \\  
1394 &  0.30$\pm$0.00 &  1.15$\pm$0.03 &  0.57 &  -83.4 &  18.16 &  ... &  ... &  2.14 \\  
1533 &  0.33$\pm$0.03 &  1.21$\pm$0.25 &  0.91 &  -82.3 &  20.11 &  ... &  ... &  0.66 \\  
1666 &  0.61$\pm$0.01 &  0.57$\pm$0.01 &  0.63 &  5.7 &  18.35 &  1.76 &  0.35 &  2.36 \\  
1768 &  0.14$\pm$0.01 &  2.05$\pm$0.16 &  0.60 &  -78.6 &  16.90 &  ... &  ... &  2.10 \\  
1826 &  0.35$\pm$0.03 &  3.95$\pm$0.28 &  0.69 &  15.2 &  19.30 &  1.993 &  0.40 &  1.93 \\  
1942 &  0.31$\pm$0.01 &  1.99$\pm$0.26 &  0.91 &  -4.1 &  18.02 &  ... &  ... &  2.05 \\  
2066 &  0.21$\pm$0.01 &  1.62$\pm$0.17 &  0.70 &  46.6 &  17.71 &  2.91 &  0.35 &  2.02 \\  
2083 &  0.12$\pm$0.00 &  4.04$\pm$0.18 &  0.53 &  -3.0 &  16.07 &  ... &  ... &  1.93 \\  
2049 &  0.17$\pm$0.02 &  2.45$\pm$0.50 &  0.54 &  -75.3 &  17.61 &  ... &  ... &  2.44 \\  
2282 &  0.57$\pm$0.03 &  0.76$\pm$0.06 &  0.59 &  80.1 &  20.20 &  ... &  ... &  2.11 \\  
2411 &  0.54$\pm$0.01 &  0.56$\pm$0.03 &  0.74 &  -21.9 &  19.67 &  ... &  ... &  2.18 \\  
2564 &  0.11$\pm$0.00 &  1.69$\pm$0.07 &  0.47 &  -27.9 &  16.31 &  ... &  ... &  1.80 \\  
2734 &  0.21$\pm$0.01 &  1.60$\pm$0.21 &  0.77 &  -14.5 &  18.54 &  ... &  ... &  2.11 \\  
2678 &  0.11$\pm$0.00 &  2.39$\pm$0.09 &  0.35 &  -74.6 &  15.17 &  ... &  ... &  2.06 \\  
2764 &  0.29$\pm$0.01 &  1.01$\pm$0.09 &  0.84 &  -52.7 &  18.85 &  ... &  ... &  1.98 \\  
2902 &  0.22$\pm$0.01 &  1.88$\pm$0.18 &  0.50 &  -15.7 &  18.46 &  ... &  ... &  2.02 \\  
2837 &  0.35$\pm$0.00 &  0.35$\pm$0.02 &  0.42 &  12.3 &  17.92 &  1.79 &  0.30 &  1.72 \\  
2965 &  0.15$\pm$0.00 &  1.70$\pm$0.11 &  0.62 &  -57.9 &  16.74 &  ... &  ... &  3.03 \\  
3036 &  0.20$\pm$0.00 &  2.15$\pm$0.10 &  0.50 &  -31.9 &  17.41 &  ... &  ... &  1.66 \\  
3126 &  0.36$\pm$0.01 &  1.27$\pm$0.07 &  0.57 &  62.4 &  19.35 &  ... &  ... &  1.84 \\  
3250 &  0.34$\pm$0.01 &  0.34$\pm$0.04 &  0.27 &  27.9 &  17.71 &  ... &  ... &  2.10 \\  
3422 &  0.35$\pm$0.02 &  0.50$\pm$0.00 &  0.37 &  -64.4 &  18.13 &  ... &  ... &  2.16 \\  
3387 &  0.05$\pm$0.00 &  2.50$\pm$0.00 &  0.63 &  -57.8 &  15.65 &  ... &  ... &  1.76 \\  
3582 &  0.27$\pm$0.02 &  1.92$\pm$0.37 &  0.77 &  74.6 &  19.33 &  ... &  ... &  2.34 \\  
3629 &  0.19$\pm$0.00 &  1.40$\pm$0.04 &  0.70 &  47.5 &  16.94 &  ... &  ... &  2.16 \\  
3818 &  0.40$\pm$0.00 &  2.07$\pm$0.03 &  0.79 &  -21.4 &  18.98 &  ... &  ... &  1.62 \\  
3766 &  0.28$\pm$0.00 &  1.79$\pm$0.06 &  0.77 &  -40.1 &  18.23 &  ... &  ... &  2.12 \\  
3822 &  0.21$\pm$0.01 &  2.33$\pm$0.23 &  0.84 &  -45.3 &  19.04 &  ... &  ... &  2.33 \\  
3970 &  0.26$\pm$0.02 &  3.67$\pm$0.36 &  0.59 &  -5.1 &  18.60 &  ... &  ... &  2.17 \\  
4121 &  0.06$\pm$0.00 &  2.20$\pm$0.00 &  0.34 &  -74.3 &  14.17 &  ... &  ... &  4.38 \\  
4033 &  0.54$\pm$0.01 &  1.28$\pm$0.02 &  0.54 &  -2.6 &  18.65 &  ... &  ... &  1.92 \\  
4239 &  0.38$\pm$0.02 &  1.20$\pm$0.09 &  0.52 &  -42.5 &  19.05 &  ... &  ... &  1.87 \\  
\hline
\end{tabular} \\
Notes. (a) The values of the errors on $R_{\rm e}$ and Sersic $n$ are representative of the 
1 $\sigma$ model errors
from GALFIT (see Buitrago et al. 2008).  This does not take into account many possible
sources of error that may bias these measurements, including magnitude of galaxy, concentration
of its light profile, etc.  The uncertainty in these structural parameters increase by on
order of 10 percent for $R_{\rm e}$ and 20 percent for $n$ due to changes in the PSF across
the NICMOS NIC3 field of view.  Also listed is the fitted axis ratios for these galaxies (b/a), and
position angles (P.A.).  The effective surface brightness ($\mu_{\rm eff}$) is listed, as is
the spectroscopic redshift ($z_{\rm spec}$), if available.  The value of `sep' is the difference
between the position of an object and the identification of the spectroscopic target, in arcsec.
Finally, the calculated photometric redshift, $z_{\rm phot}$, is shown.
\end{minipage}
\end{table*}

\vspace{1cm}
\setcounter{table}{5}
\begin{table*}
\begin{minipage}{160mm}
 \caption{Structural parameters for galaxies with $M_{*} > 10^{11}$ \solm within the GOODS-S field.}
 \label{tab1}
 \begin{tabular}{@{}cccccccccc}
 \hline
\hline
ID & $R{\rm e}$ (arcsec) & Sersic $n$ & b/a & P.A. (deg) & $\mu_{\rm eff}$ (mag arcsec$^{-2}$) & $z_{\rm spec}$ & sep (arcsec) & $z_{\rm phot}$ \\
\hline 
4299 &  0.31$\pm$0.01 &  0.65$\pm$0.06 &  0.46 &  -8.0 &  18.57 &  ... &  ... &  0.31 \\  
4281 &  0.20$\pm$0.02 &  3.03$\pm$0.57 &  0.57 &  -39.8 &  18.61 &  ... &  ... &  1.70 \\  
4348 &  0.34$\pm$0.02 &  1.78$\pm$0.17 &  0.58 &  85.3 &  19.84 &  ... &  ... &  0.76 \\  
4434 &  0.13$\pm$0.00 &  1.35$\pm$0.05 &  0.67 &  -71.4 &  16.56 &  2.09 &  0.12 &  1.96 \\  
4399 &  0.08$\pm$0.01 &  3.40$\pm$1.00 &  0.70 &  -69.9 &  17.70 &  ... &  ... &  2.31 \\  
4557 &  0.34$\pm$0.01 &  1.48$\pm$0.05 &  0.38 &  72.6 &  17.45 &  ... &  ... &  1.75 \\  
4754 &  0.38$\pm$0.01 &  0.75$\pm$0.03 &  0.44 &  -37.6 &  18.68 &  ... &  ... &  2.21 \\  
4706 &  0.06$\pm$0.00 &  4.62$\pm$0.19 &  0.50 &  -75.7 &  14.69 &  2.34 &  0.08 &  1.77 \\  
4882 &  0.10$\pm$0.00 &  2.25$\pm$0.08 &  0.65 &  2.8 &  16.44 &  ... &  ... &  1.41 \\  
4941 &  0.24$\pm$0.16 &  3.79$\pm$4.42 &  0.80 &  -22.8 &  18.42 &  ... &  ... &  2.87 \\  
5171 &  0.10$\pm$0.00 &  3.36$\pm$0.28 &  0.66 &  -32.0 &  16.50 &  2.40 &  0.21 &  2.58 \\  
5281 &  0.48$\pm$0.01 &  1.14$\pm$0.02 &  0.70 &  16.1 &  19.03 &  ... &  ... &  0.69 \\  
5445 &  0.42$\pm$0.03 &  1.06$\pm$0.08 &  0.52 &  61.7 &  19.36 &  ... &  ... &  2.39 \\  
5372 &  0.20$\pm$0.02 &  1.64$\pm$0.22 &  0.42 &  -32.2 &  17.63 &  ... &  ... &  2.53 \\  
5533 &  0.09$\pm$0.00 &  3.20$\pm$0.32 &  0.63 &  -33.5 &  16.29 &  ... &  ... &  2.37 \\  
5524 &  0.06$\pm$0.00 &  1.00$\pm$0.06 &  0.19 &  -22.8 &  12.48 &  ... &  ... &  2.82 \\  
5764 &  0.16$\pm$0.01 &  2.88$\pm$0.15 &  0.64 &  -27.6 &  16.95 &  ... &  ... &  2.34 \\  
5853 &  0.24$\pm$0.01 &  0.38$\pm$0.05 &  0.49 &  -84.0 &  17.78 &  2.41 &  0.16 &  1.85 \\  
5933 &  0.27$\pm$0.00 &  1.08$\pm$0.05 &  0.92 &  64.0 &  18.84 &  ... &  ... &  2.48 \\  
6035 &  0.22$\pm$0.01 &  4.27$\pm$0.19 &  0.83 &  -67.9 &  18.19 &  ... &  ... &  2.75 \\  
6114 &  0.11$\pm$0.00 &  3.90$\pm$0.12 &  0.72 &  -68.4 &  16.00 &  ... &  ... &  1.55 \\  
6220 &  0.40$\pm$0.01 &  1.45$\pm$0.05 &  0.80 &  -73.5 &  19.57 &  ... &  ... &  1.55 \\  
6352 &  0.50$\pm$0.03 &  2.59$\pm$0.11 &  0.51 &  65.7 &  19.54 &  ... &  ... &  2.56 \\  
6468 &  0.31$\pm$0.01 &  0.98$\pm$0.04 &  0.46 &  -44.8 &  17.45 &  ... &  ... &  1.76 \\  
6584 &  0.34$\pm$0.01 &  1.10$\pm$0.03 &  0.22 &  68.8 &  17.16 &  ... &  ... &  1.66 \\  
6575 &  0.41$\pm$0.01 &  0.33$\pm$0.02 &  0.73 &  -62.6 &  18.30 &  ... &  ... &  1.52 \\  
6876 &  0.29$\pm$0.01 &  0.47$\pm$0.04 &  0.35 &  -82.7 &  17.18 &  ... &  ... &  1.35 \\  
7090 &  0.37$\pm$0.01 &  2.68$\pm$0.06 &  0.50 &  -65.7 &  17.89 &  ... &  ... &  2.58 \\  
7155 &  0.10$\pm$0.01 &  4.52$\pm$0.35 &  0.51 &  -66.8 &  15.67 &  ... &  ... &  0.49 \\  
7320 &  0.12$\pm$0.00 &  1.12$\pm$0.05 &  0.56 &  -31.6 &  16.18 &  ... &  ... &  1.39 \\  
7425 &  0.16$\pm$0.00 &  2.13$\pm$0.06 &  0.45 &  66.4 &  16.48 &  1.31 &  0.2 &  1.40 \\  
7677 &  0.21$\pm$0.00 &  2.85$\pm$0.09 &  0.83 &  50.0 &  17.01 &  1.19 &  0.15 &  1.34 \\  
7970 &  0.37$\pm$0.01 &  0.56$\pm$0.02 &  0.84 &  -79.4 &  18.73 &  ... &  ... &  1.72 \\  
8140 &  0.19$\pm$0.00 &  2.72$\pm$0.07 &  0.77 &  -19.8 &  17.24 &  1.90 &  0.14 &  1.78 \\  
8213 &  0.15$\pm$0.00 &  1.69$\pm$0.07 &  0.63 &  53.8 &  16.56 &  ... &  ... &  0.70 \\  
\hline
\end{tabular} \\
Notes. (a) The values of the errors on $R_{\rm e}$ and Sersic $n$ are representative of the 
1 $\sigma$ model errors
from GALFIT (see Buitrago et al. 2008).  This does not take into account many possible
sources of error that may bias these measurements, including magnitude of galaxy, concentration
of its light profile, etc.  The uncertainty in these structural parameters increase by on
order of 10 percent for $R_{\rm e}$ and 20 percent for $n$ due to changes in the PSF across
the NICMOS NIC3 field of view.    Also listed is the fitted axis ratios for these galaxies (b/a), and
position angles (P.A.).  The effective surface brightness ($\mu_{\rm eff}$) is listed, as is
the spectroscopic redshift ($z_{\rm spec}$), if available.  The value of `sep' is the difference
between the position of an object and the identification of the spectroscopic target, in arcsec.
Finally, the calculated photometric redshift, $z_{\rm phot}$, is shown.
\end{minipage}
\end{table*}

\vspace{-0.5cm}

\appendix

\label{lastpage}

\end{document}